\documentclass{article}
\usepackage[utf8]{inputenc}
\usepackage[a4paper, total={6in, 9in}]{geometry}

\usepackage{cite}
\usepackage{amsmath,amssymb,amsfonts,bm}
\usepackage{algorithmic}
\usepackage{graphicx}
\usepackage{textcomp}
\usepackage{xcolor}
\usepackage{todonotes}
\usepackage{multirow}
\usepackage{tabu}
\usepackage{enumerate}
\usepackage{float}
\usepackage{enumitem}
\usepackage{caption}
\usepackage{subcaption}
\usepackage{appendix}
\usepackage{algorithm}
\usepackage{booktabs}
\usepackage{comment}
\usepackage{titlesec}
\usepackage{amsmath}               
  {
      
  }

\DeclareMathOperator*{\Var}{Var} 
\DeclareMathOperator*{\sign}{sign}

\DeclareMathOperator*{\argmin}{argmin}

\title{Block-diagonal idiosyncratic covariance estimation in high-dimensional factor models for financial time series}

\date{June 2024}

\begin{document}

\author{ Lucija Žignić$^*$, Stjepan Begušić$^\dagger$, and Zvonko Kostanjčar$^\dagger$ 
\vspace{10pt}
\\ \small
$^*${Forvis Mazars, Management Consulting, Strojarska 20, 10000 Zagreb, Croatia}\\ \small
$^\dagger${University of Zagreb, Faculty of Electrical Engineering and Computing, Unska 3, 10000 Zagreb, Croatia}\\ 
\small
Email addresses: Lucija.Zignic@fer.unizg.hr\footnote{(corresponding author)}, Stjepan.Begusic@fer.unizg.hr, Zvonko.Kostanjcar@fer.unizg.hr
}

\maketitle

\begin{abstract}
Estimation of high-dimensional covariance matrices in latent factor models is an important topic in many fields and especially in finance. Since the number of financial assets grows while the estimation window length remains of limited size, the often used sample estimator yields noisy estimates which are not even positive definite. Under the assumption of latent factor models, the covariance matrix is decomposed into a common low-rank component and a full-rank idiosyncratic component. In this paper we focus on the estimation of the idiosyncratic component, under the assumption of a grouped structure of the time series, which may arise due to specific factors such as industries, asset classes or countries. We propose a generalized methodology for estimation of the block-diagonal idiosyncratic component by clustering the residual series and applying shrinkage to the obtained blocks in order to ensure positive definiteness. We derive two different estimators based on different clustering methods and test their performance using simulation and historical data. The proposed methods are shown to provide reliable estimates and outperform other state-of-the-art estimators based on thresholding methods.
\end{abstract}

\vspace{5mm} 


\section{Introduction}
Covariance matrix estimation is a heavily researched topic in many fields, and is a crucial component for risk modeling in finance, where risk models rely on the estimation of the asset return covariance \cite{LedoitPower2022,Lam2020b,Choi2019}. With the growth of the number of financial assets, high dimensionality of these estimates becomes an issue -- the sample estimates may be noise driven and no more reliable \cite{Bun2017,MohsenPourahmadi2013}. Moreover, due to the dynamic nature of financial markets, estimates from long historical data may be obsolete and relatively short time windows are used instead -- this setting of high dimension and low sample size (where the number of variables $p$ exceeds the sample size $n$) is very common in finance today \cite{Goldberg2022}. Fortunately, financial markets also display a certain level of structure which can be used to obtain reliable estimates in such adverse environments. Mainly, asset pricing literature finds that a sizeable amount of variance in large panels of asset return data is driven by a smaller number of factors \cite{Ross1976,Fama1993,Fama2015}. Asset return dynamics and their correlations are thus often explained using factor models, with a common component (from exposure to these common factors), and an idiosyncratic component (specific for each asset) \cite{Connor1995,Fan2011a}. Under some reasonable assumptions, the asset return covariance under such a model is the sum of a common covariance component (which is low-rank, since the number of factors is much lower than the number of assets) and an idiosyncratic covariance component. This leads to a number of structured and well-conditioned estimators of the covariance matrix which mostly amount to estimating the factor model parameters \cite{Fan2011a, Fan2013}. However, these estimators focus mostly on the identification of pervasive factors, their interpretation, and performance in asset pricing \cite{Ross1976, Connor1995, Lettau2020}. The correlation structures within the idiosyncratic components have received comparatively little attention. Since these correlations are likely due to exposure to non-pervasive factors such as sectors, countries, or asset classes, ignoring these factors considerably reduces the performance of the estimators \cite{Fan2016}. 

In this paper we focus on the problem of structured estimation of the idiosyncratic covariance component in high-dimensional factor models, based on the assumption that the idiosyncratic correlations arise between assets exposed to some common but non-pervasive factors \cite{Fan2016,Begusic2019}. An important requirement is to ensure positive-definite estimates of covariance matrix estimates even in the high-dimension-low-sample-size setting of $p>n$. Some of the early approaches based on high-dimensional factor models ensured positive definiteness by assuming a diagonal idiosyncratic covariance \cite{Fan2008}, which completely ignores the elements of risk arising from the correlations between the idiosyncratic components. More appropriately, assuming sparsity of the off-diagonal correlations in the idiosyncratic components allows for the approximate factor model structure \cite{Chamberlain1983}. A number of thresholding procedures have been devised earlier with the goal of estimating sparse covariance matrices (assuming sparsity of the entire covariance) \cite{Bickel2008,Rothman2009,Cai2011}. Combining the high-dimensional factor structure and the assumption of sparsity of the idiosyncratic covariance (i.e. conditional sparsity) led to estimators such as the POET \cite{Fan2013} and S-POET \cite{Wang2017}, which were shown to produce estimates which perform well for some portfolio optimization use cases. However, the thresholding methods used in these estimators do not exploit any common structures in the idiosyncratic components, which are known to occur due to sector, asset class or other non-pervasive factor exposure \cite{Fan2016,Ait-Sahalia2017,Begusic2020}. In this paper we use these structures to our advantage: by assuming that the idiosyncratic components exhibit correlations due to some group-specific factors such as asset class or sector classification, their covariance structure becomes block-diagonal. This leads to a potentially wider set of positive definite estimates, and allows for a richer description \cite{Zignic2022}. As the main information is extracted in the common component, the factors that may exist within the clusters will generally be weak and hard to identify. Moreover, the unknown cluster membership together with the unknown number of factors within each cluster additionally complicates estimation procedures \cite{Ando2017, Begusic2020}. Finally, the groupings themselves may not be easily incorporated into linear factor models if the effects of the cluster-specific sources of variation are not linear. To allow for the latter and avoid any formerly mentioned obstacles, we focus on treating the cluster-specific dependencies as the idiosyncratic component of the covariance in the factor model. 

We develop a set of estimators which firstly calculate the low-rank common covariance component using principal components, and then use the residuals to estimate the unknown group memberships and the resulting block-diagonal idiosyncratic covariance. We use several clustering approaches to estimate these groups, and propose a cross-validation procedure for selecting the optimal grouping (and consequently the idiosyncratic covariance). Since the cluster sizes are allowed to grow beyond the sample size, we also apply covariance shrinkage to each of the blocks to ensure positive definiteness of the estimates even in high-dimension-low-sample-size settings. This allows us to conduct a comprehensive study of the performance of different covariance estimators for high-dimensional factor models with a block-diagonal idiosyncratic covariance structure. We develop a simulation framework to test various settings and configurations of these blocks, and also apply the developed estimators to historical market data. To measure the performance of the estimates we consider both the measures of how well the idiosyncratic covariance patterns are identified, and the out-of-sample performance of portfolios constructed using the covariance estimates. Simulations show that the estimation method using a hierarchical clustering approach is able to perform very good in both sparse performance measures and overall, and is able to successfully estimate structures with very small clusters. Results on historical data show excellent out of sample performance of all the clustering approaches and concludes with the observed differences of the final idiosyncratic covariance estimates obtained by the different clustering approaches.

\section{Model}
Let $Y$ denote the $p$-dimensional random vector of asset returns\footnote{We consider arithmetic returns $Y_{t} = R_{t}/R_{t-1}-1$, where $R_{t}$ is the financial asset price at time step $t$. Arithmetic returns allow for efficient matrix operations to be used in portfolio return calculations, which leads to simple calculations for the portfolio variance which is simply the variance of a linear combination of asset returns. For more details see \cite{Dorfleitner2003}.}
for $p$ assets. We consider the latent factor model
	\begin{equation}
	Y_{}= \mathbf{B}F + \varepsilon_{},
		\label{eq:factor_model}
	\end{equation}
where $\mathbf{B}$ is a $p \times K$ matrix of factor loadings, $F_{}$ is a $K$-dimensional random vector of $K$ common factors and $\varepsilon_{}$ is a $p$-dimensional random vector of the specific factors, also known as the idiosyncratic component. The factor loadings $\mathbf{B}$, the factor realizations $F$ and the idiosyncratic component $\varepsilon$ are considered to be unobservable, so the factor model parameters need to be estimated from the observable asset returns. The idiosyncratic and common factors are assumed to be uncorrelated \cite{MohsenPourahmadi2013}, which is a common assumption helping with their identifiability. Under the model, the asset return covariance $\mathrm{Cov}(Y) =\mathbf{\Sigma}$ has the following decomposition \cite{Fan2011a}:
\begin{equation}
	\mathbf{\Sigma} = \mathbf{B} \mathrm{Cov}(F) \mathbf{B'} + \mathbf{\Psi}, 
\end{equation}
where $\Psi = \mathrm{Cov}(\varepsilon)$ is the covariance of the specific factors, also known as the idiosyncratic covariance. The common covariance component $\mathbf{B} \mathrm{Cov}(F) \mathbf{B'}$ is low-rank (since $K < p$), and explains the majority of the correlations between different assets as the result of their exposure to a smaller number of common factors. The idiosyncratic covariance $\mathbf{\Psi}$ is often considered to be diagonal, however in this paper we consider models from the category of approximate factor models \cite{Chamberlain1983}, where some sparse correlations between idiosyncratic components are allowed -- thus the idiosyncratic covariance is full rank and sparse. Moreover, motivated by the documented grouping of financial assets (within industries or asset classes) \cite{Fan2016,Begusic2019,Begusic2020}, we allow the idiosyncratic components of asset returns to be associated with one of a total of $M$ clusters. The idiosyncratic components between assets within the same group may be correlated, and the idiosyncratic components of asset pairs from different groups are uncorrelated. This means that (if the assets were sorted according to group membership) the idiosyncratic covariance has a block-diagonal structure. Note this setting does not exclude singleton clusters with only one asset whose idiosyncratic component is uncorrelated to all the others.

Let $c_m$ denote the subset of assets within cluster $m$ (where $m \in {1,...,M}$). If the assets are sorted according to their cluster membership then the idiosyncratic covariance matrix has the following block-diagonal structure: 
\begin{equation}
	\begin{aligned}
		&\mathbf{\Psi} =
		\begin{bmatrix}
			\mathbf{\Psi}^{(c_1)} & 0 & \ldots & 0 \\
			0 & 	\mathbf{\Psi}^{(c_2)} & \ldots & 0 \\
			\vdots & \vdots & \ddots & \vdots \\
			0 & 0 & \ldots & 	\mathbf{\Psi}^{(c_M)} \\
		\end{bmatrix},
	\end{aligned} 
\end{equation}
where $\mathbf{\Psi}^{(c_m)}$ is the idiosyncratic covariance of all assets within cluster $m$. We allow for the setting where the number of clusters $M$ is large, even as large as or close to $p$ (meaning that all assets belong to their own cluster, and that the resulting covariance matrix is diagonal). The cluster memberships, the number of clusters and their sizes are all considered unknown and need to be estimated from the data. Ultimately, once these are known, the covariance elements themselves need to be estimated as well.

To ensure the identifiability of the factors and factor loadings in the latent factor model, a usual restriction is that $\mathbf{B}'\mathbf{B}$ is diagonal and $\mathrm{Cov}(F) = \mathbf{I}_p$ \cite{Fan2013}. In order to be able to estimate and distinguish the factors from the idiosyncratic components, several assumptions are imposed on the spectrum of the asset return covariance. Firstly, the eigenvalues associated with the common factors (the largest $K$ eigenvalues of $\mathbf{\Sigma}$) are unbounded and assumed to grow with growing dimensionality $p$. Secondly, the eigenvalues of the idiosyncratic covariance are bounded as $p$ grows, so that they do not "leak" into the spectrum of the common component. This enables a two-step estimation approach such as in the POET and S-POET estimators \cite{Fan2013,Wang2017}, which we also follow in this paper.

\section{Estimation}	
All of the model parameters -- the factor loadings and idiosyncratic covariance -- need to be estimated from the data sample $\mathbf{Y} \in \mathbb{R}^{p \times T}$. In this section and the rest of the paper, we refer to the estimates of the asset return covariance as ${\widehat{\mathbf{\Sigma}}}$, with the index noting the type of estimator. The primary important estimator is the sample covariance:
\begin{equation}
    {\widehat{\mathbf{\Sigma}}_s} = \frac{1}{T-1}\sum_{t = 1}^{T}(Y_t - \overline Y)(Y_t - \overline Y)',
\end{equation}
where $Y_t$ is the $p$-dimensional vector of asset returns at time $t$ and $\overline Y$ is the $p$-dimensional sample mean return. Following the assumptions stated in the previous section, the $p \times K$ matrix of factor loadings $\mathbf{B}$ can be estimated as $\widehat{\mathbf{B}} = [\sqrt{\widehat{\lambda}_1}\widehat\Gamma_1,...,\sqrt{\widehat{\lambda}_K}\widehat\Gamma_K]$, where $\widehat{\lambda}_1 \geq \widehat{\lambda}_2 \geq ... \geq \widehat{\lambda}_p$ are the eigenvalues and ${\widehat{\Gamma}}_i$, $i= 1,...p$ the corresponding eigenvectors of the sample covariance matrix ${\widehat{\mathbf{\Sigma}}}_{s}$ \cite{Bai2019}. Thus, the estimators considered and proposed in this paper are of the following form: 
	\begin{align}
		{\widehat{\mathbf{\Sigma}}} = \sum_{i=1}^{K} \widehat{\lambda}_i {\widehat{\Gamma}}_i {\widehat{\Gamma}}^{'}_i + {\widehat{\mathbf{\Psi}}},
		\label{eq:poet}
	\end{align}
where $\widehat{\mathbf{\Psi}}$ is the estimate of the idiosyncratic covariance matrix. 
Recent results on the asymptotics of the eigenstructure of high-dimensional covariance matrices suggest that the eigenvalue estimates are biased \cite{Wang2017}. To mitigate this estimation bias, we replace the sample estimates $\widehat{\lambda}_i$ in the estimator \eqref{eq:poet} with the shrunk eigenvalues:
\begin{equation}
    \widehat{\lambda}^{S}_{i} = \max \{\widehat{\lambda}_{i} - c p / T, 0\},
    \label{eq:SPOET}
\end{equation} 
where $c$ is calculated as:
\begin{equation}
    \widehat{c} = \Big (\text{tr}({\widehat{\mathbf{\Sigma}}_s}) - \dfrac{\sum_{i=1}^{K} \widehat{\lambda}_{i}}{(p-K-p K / T) }\Big ).
\end{equation}
Note that the bias correction term $c p / T$ in \eqref{eq:SPOET} diminishes as the number of samples $T$ grows with respect to the dimensionality $p$. Since we deal with high-dimensional cases when $p>T$, this term will not be negligible. 

The estimators generally follow a two-step procedure:
\begin{enumerate}
    \item Estimate the common component $\sum_{i=1}^{K} \widehat{\lambda}^S_i {\widehat{\Gamma}}_i {\widehat{\Gamma}}^{'}_i$, using the first $K$ principal components: $\widehat{\lambda}^S_i$ are the shrunk eigenvalues from \eqref{eq:SPOET} and $\widehat{\Gamma}^{'}_i$ are the corresponding sample eigenvectors.
    \item Apply a sparse estimation procedure to the residual covariance matrix, also known as the orthogonal complement: $\widehat{\mathbf{S}} = {\widehat{\mathbf{\Sigma}}}_{s} - \sum_{i=1}^{K} \widehat{\lambda}^S_i {\widehat{\Gamma}}_i {\widehat{\Gamma}}^{'}_i$, in order to obtain a sparse estimate of $\mathbf{\Psi}$.
\end{enumerate}
It is important to note that the orthogonal complement $\widehat{\mathbf{S}}$ is a full matrix of rank $\mathrm{min}(n,T)-K$ which does not serve as an idiosyncratic covariance estimate $\widehat{\mathbf{\Psi}}$ (since it is not a sparse matrix, nor a full-rank matrix). Different estimates $\widehat{\mathbf{\Psi}}$ are obtained from $\widehat{\mathbf{S}}$ by applying some sparsity-inducing procedures (such as thresholding or the proposed clustering based estimation). 
The first step described above is based on the sample principal components, and is common to all of the estimators considered in this paper. What this paper focuses on is the second step -- the idiosyncratic covariance estimation, in the presence of clustered specific components, with an unknown clustering. The following sections lay out the elements of the estimation procedures for different important quantities.

\subsection{Estimating the number of factors}
\label{subsubsec:fac}
An important issue to deal with before we delve deeply into the specific of the estimators is the estimation of the number of factors $K$. In this paper we follow the Bai-Ng approach and use an information criterion (labeled $IC1$ in the original paper \cite{Bai2002}). The Bai-Ng information criterion ($IC$) defines the procedure to estimate $K$ as
\begin{equation}
    \begin{split}
\widehat{K} = & \argmin_{0 \leq \tilde K \leq N} \log \bigg \{ \dfrac{1}{pT} \| \mathbf{Y} - \mathbf{Y} {\widehat{\mathbf{B}}} \mathbf{\Delta}^{-1} {\widehat{\mathbf{B}}}' \|_F^2 \bigg\} \\ &
 + \tilde K \bigg ( \dfrac{p+T}{pT} \log{\bigg( \dfrac{p+T}{pT} \bigg)} \bigg ), 
\end{split}
\label{eq:BaiNgIC}
	\end{equation}
where $N$ is an upper bound for the possible number of latent factors (often set to $\text{min}(T,p)$), ${\widehat{B}}$ is the $p \times \tilde K$ loadings matrix estimate for $\tilde K$ factors, and $\mathbf{\Delta}$ is a $\tilde K \times \tilde K$ diagonal matrix with the $\tilde K$ largest eigenvalue estimates on the diagonal. 
The first term in Equation \ref{eq:BaiNgIC} describes the log mean square error of reconstructing the original data sample $\mathbf{Y}$ using the estimated factor model, which is reduced by increasing the number of factors. The second term is a penalization term which grows with the number of considered factors $\tilde K$. Ultimately, the information criterion balances the reduced reconstruction error with the added complexity of the model and will result with an estimate of the number of factors $\hat K$ which yields the best reduction in error for the smallest number of factors.

This procedure yields in choosing first $\hat K$ eigenvalues which have significantly higher value than the rest, thus making it worth to be established as factors. The rest of the eigenvalues, with much lower amount of carrying information are thus left in the orthogonal complement matrix and subject to sparsity inducing methods.

\subsection{Estimating the idiosyncratic covariance via thresholding}
\label{thresholding}
The state-of-the-art estimators most commonly use generalized thresholding procedures \cite{Rothman2009,Fan2013,Ando2017,Begusic2020}, the resulting sparse estimates of $\mathbf{\Psi}$ have no underlying structure, and are limited to a very narrow range of possible estimates which are positive-definite. The most sophisticated thresholding methods include adaptive thresholding, applied to the orthogonal complement matrix $\widehat{\mathbf{S}} = (\widehat{S}_{ij})$. The idea is to apply generalized thresholding operator function $f_{\tau_{ij}}$ to the full covariance ${\widehat{\mathbf{S}}}$ in order to obtain the sparse estimate ${\widehat{\mathbf{\Psi}}_\tau}$:
\begin{equation} \label{adaptive}
	{\widehat{\mathbf{\Psi}}}_{\tau_{ij}} =
	\begin{cases}
		\widehat{S}_{ii}  & i=j\\
		f_{\tau_{ij}} ( \widehat{S}_{ij})  & i \neq j.
	\end{cases}        
\end{equation}
For any $\tau_{ij} \geq 0$, the generalized thresholding operator is a function $f_{\tau_{ij}}: \mathbb{R} \rightarrow \mathbb{R}$ which, for all $z \in \mathbb{R}$ satisfies the following conditions: \cite{Rothman2009}:
\begin{enumerate}
	\item $|f_{\tau_{ij}}(z)| \leq |z|$,
	\item  $f_{\tau_{ij}}(z) = 0$ for $|z| \leq \tau_{ij}$,
	\item $|f_{\tau_{ij}} (z) - z| \leq \tau_{ij}$.
\end{enumerate}
These conditions are satisfied by several popular thresholding functions, out of which we consider the following:
\begin{itemize}
	\item hard thresholding \label{hard}: 
	\begin{equation}
		s_\tau^{HT} (z) = z \mathbf{1}_{(|s_{ij}| \geq \tau_{ij})},
	\end{equation}
	\item soft thresholding \cite{Rothman2009}: 
		\begin{equation}
		f_\tau^{ST} (z) = \sign(z)(\vert z\vert -\tau_{ij})_{+} ,
	\end{equation}
	\item adaptive lasso \cite{Zou2006}:
		\begin{equation}
		f_\tau^{AL} (z) = \sign(z)(|z|-\tau_{ij}^{a+1}|z|^{-a})_{+} ,
	\end{equation}
	\item SCAD (smoothly clipped absolute deviation) \cite{Fan2001}:
	\end{itemize}
	\begin{equation} 
	f_\tau^{SCAD} (z) =
		\begin{cases}
		\sign(z)(\vert z\vert -\tau_{ij})_{+},   & |z| \leq 2 \tau_{ij}\\
\dfrac{[(a-1)z-\sign(z) a \tau_{ij}]}{(a-2)},	   & 2 \tau_{ij} < |z| \leq a \tau_{ij} \\
	z, & |z| > a \tau_{ij}.
		\end{cases}        
	\end{equation}
The adaptive thresholding parameter \cite{Cai2011} is of the form
\begin{equation} \label{para}
	\tau_{ij} = \tau \sqrt{\dfrac{\widehat{\theta}_{ij} \log p}{T}},
\end{equation}
where $\tau$ is a tuning parameter and $\widehat{\theta}_{ij}$ are estimates of $\theta_{ij}=\Var [(Y_i-\mu_i) (Y_j-\mu_j)]$. 
The parameter $\tau$ can be set as fixed or obtained through a data-driven cross-validation procedure \cite{Cai2011}. In the latter case, the procedure iterates over the space of possible values of $\tau$ for which the estimates $\widehat{\mathbf{\Psi}}$ are positive-definite. On the one side, for large values of $\tau$ the estimates become diagonal. Unfortunately, for lower values of $\tau$ (which allow more non-zero entries) the matrices quickly stop being positive-definite (due to the non-zero entries not following any specific patterns), thus narrowing the space of acceptable values of $\tau$ and limiting the estimators \cite{Fan2013}.

Through the paper, \textit{thresholding based} estimators are denoted with $\widehat{\mathbf{\Psi}}_{SOFT}$ (SOFT) for the soft thresholding function, $\widehat{\mathbf{\Psi}}_{AL}$ (AL) for the adaptive lasso thresholding function, and $\widehat{\mathbf{\Psi}}_{SCAD}$ (SCAD) for the SCAD thresholding function. 

As mentioned, these estimators provide sparse estimates with no inherent structure, thus potentially missing out on certain narrow factors such as sectors, asset classes or countries. Recent literature has documented that financial assets exhibit clustering patterns, even when the common factors are filtered out \cite{Ait-Sahalia2017,Fan2016a,Begusic2020}. This motivates the approach proposed in this paper -- under the hypothesis that the idiosyncratic covariance reflect this grouping in the residual time series, we formulate a number of \textit{clustering based} estimators.

\subsection{Block-diagonal idiosyncratic covariance estimation} 
We denote the group membership information as a zero-one $p \times p$ indicator matrix $\mathbf{C}$ (also known as a \textit{mask}), where the element $C_{ij} = 1$ if $i$ and $j$ are in the same cluster group $c_m$, for $i,j \in {1,...,p}$. If the rows and columns of $\mathbf{C}$ (and consequently, the $p$ assets in the factor model \eqref{eq:factor_model}) are sorted according to their cluster membership, then $\mathbf{C}$ is a block-diagonal matrix. Without loss of generality, in the following notation we assume that the assets are sorted according to their cluster membership (this can also be done once the clustering is known) and that $\mathbf{C}$ is block-diagonal.

Let matrix $\mathbf{C}$ contain $\mathbf{C}_1$, $\mathbf{C}_2$,...,$\mathbf{C}_M$ cluster blocks for each of the $M$ clusters. The imposed block-diagonal idiosyncratic covariance $\mathbf{\widehat{\Psi}}^{C}$ which is obtained from the orthogonal complement $\widehat{\mathbf{S}}$ (the initial full idiosyncratic covariance estimate) is:
\begin{equation} \label{block}
	\begin{aligned}
		&\mathbf{\widehat{\Psi}}^{C} = (\widehat{S}_{ij} \mathbf{1}_ {(ij) \in \mathbf{C}}) 
		= \widehat{\mathbf{S}} \circ \mathbf{C} = \\ &  
		= \widehat{\mathbf{S}}\circ \begin{bmatrix}
			\mathbf{C}_1 & 0 & \ldots & 0 \\
			0 & \mathbf{C}_2 & \ldots & 0 \\
			\vdots & \vdots & \ddots & \vdots \\
			0 & 0 & \ldots & \mathbf{C}_M \\
		\end{bmatrix} = 
		\begin{bmatrix}
			\widehat{\mathbf{S}}^{C_1} & 0 & \ldots & 0 \\
			0 & 	\widehat{\mathbf{S}}^{C_2} & \ldots & 0 \\
			\vdots & \vdots & \ddots & \vdots \\
			0 & 0 & \ldots & 	\widehat{\mathbf{S}}^{C_M} \\
		\end{bmatrix},
	\end{aligned} 
\end{equation}
where each block is defined as $ (	\widehat{\mathbf{S}}^{C_m} = \widehat{S}_{ij} \mathbf{1}_ {(ij) \in \mathbf{C}_m})$, $m \in {1,...M}$, and $\circ$ denotes the Hadamard element-wise product. 

The approach proposed above is used for all the different estimators of the block-diagonal idiosyncratic covariance. However, it still does not guarantee positive-definiteness of the covariance estimates. For instance, when the dimension of a block $\mathbf{C}_m$ (the number of time series in cluster $m$) is larger than the length of the time series estimation window $T$, some eigenvalues of the block-diagonal idiosyncratic matrix estimate (and thus some of the eigenvalues of the entire covariance matrix estimate) are very close to zero (or exactly zero), resulting in the covariance matrix estimate which is not positive definite. The positive definiteness of the idiosyncratic covariance is important in applications where the inverse of the estimated covariance matrix is needed (for example, if we want to perform portfolio optimization using the covariance estimate). As our intention is to produce the method with no constraints on sample size as well as no constraints on the number of clusters (and thus cluster sizes), we incorporate a shrinkage method within the blocks which always results with a positive definite matrix. Although there are many different forms of shrinkage and possible shrinkage targets, to avoid additionally complicating the procedure we use linear shrinkage \cite{Ledoit2004,Ledoit2004a}, applied to each block $\widehat{\mathbf{S}}^{C_m}$ separately.

Linear shrinkage can be viewed as a weighted average of the variance part and bias part of the covariance estimates, where weights should optimize the bias-variance trade-off \cite{MohsenPourahmadi2013}. We treat each block $\widehat{\mathbf{S}}^{C_m}$, $m \in {1,...M}$ as a separate covariance matrix and perform the shrinkage procedure on it \cite{Zignic2022}. A common form of the estimator is a linear combination of the covariance matrix $\mathbf{\widehat{S}}^{C_m} = (\widehat{S}^m_{ij})$ and the shrinkage target matrix $\mathbf{\Tilde{S}}^{C_m}$, with sample variances $\widehat{{S}}^m_{ii} = [\widehat{S}^m_{11},\ldots,\widehat{{S}}^m_{pp}]'$ on the diagonal and covariances $\Tilde{r} \sqrt{\widehat{S}^m_{ii} \widehat{S}^m_{jj}}$ off diagonal, where $\Tilde{r}$ is the average of all sample pairwise correlations. The shrinkage estimator is defined as:
\begin{equation} \label{shri}
	\widehat{\mathbf{S}}^{C_m}_{s} = \alpha_m \widehat{\mathbf{S}}^{C_m} + (1-\alpha_m) \mathbf{\Tilde{S}}^{C_m},
\end{equation}
where $\alpha_m$ is a scalar parameter between $0$ and $1$ which we search for each block component $C_m$, $m \in {1,...,M}$. To estimate $\alpha_m$ from sample data, we follow the well-established Ledoit and Wolf \cite{Ledoit2004a} procedure which can be found in the \ref{alpha}. 
The resulting positive definite idiosyncratic covariance estimate is 
\begin{equation}
\widehat{\mathbf{\Psi}}^{C}_{s} = \begin{bmatrix}
	\widehat{\mathbf{S}}^{C_1}_{s} & 0 & \ldots & 0 \\
	0 & 	\widehat{\mathbf{S}}^{C_2}_{s} & \ldots & 0 \\
	\vdots & \vdots & \ddots & \vdots \\
	0 & 0 & \ldots & 	\widehat{\mathbf{S}}^{C_M}_{s} \\
\end{bmatrix}.
\end{equation}

Now, the estimation of the idiosyncratic component as a block-diagonal matrix rests solely on the method employed to determine the blocks themselves - the structure of $\mathbf{C}$.

\subsubsection{Estimating the blocks using predefined asset groups} 
The simplest approach, which we lay out here as a benchmark, is to use pre-determined classifications or groupings of assets to formulate the clusters in the idiosyncratic component \cite{Ait-Sahalia2017,Fan2016}. In this approach, the sparse component is obtained by setting to zero all the pairs which are not in the same group and leaving the sample values of the orthogonal complement for the entries corresponding to the pairs in the same group. We formulate the clustering-shrinkage estimator based on industry classifications $\widehat{\mathbf{\Psi}}_{CSI}$ ($CSI$) which relies on stock industry classification data to estimate the idiosyncratic component:
\begin{equation}
    C_{ij} = 
    \begin{cases}
		1  & i \text{ and } j \text{ are in the same asset group},\\
		0  & \mathrm{otherwise}.
	\end{cases}
\end{equation}
However, this approach suffers from several drawbacks. Firstly, the classification data (such as industries, asset classes or countries) are not always available for different datasets and asset universes. Secondly, the classification itself may not be optimal, since the grouping does not guarantee the highest asset return correlations within the groups. To alleviate these issues, we propose two {clustering based} methods to estimate the block-diagonal idiosyncratic covariance.

\subsection{Clustering based estimation of the idiosyncratic covariance}
A natural extension of the previous approaches based on industry classifications of stock data is to estimate the optimal groupings from the data. In this section we develop a procedure based on different clustering approaches to the orthogonal complement $\widehat{\mathbf{S}}$, resulting in block-diagonal estimates $\widehat{\mathbf{\Psi}}$ of the idiosyncratic covariance. The clustering procedures are applied to the residual series $\widehat{\mathbf{E}} \in \mathbb{R}^{p \times T}$:
\begin{equation}
    \widehat{\mathbf{E}} = \mathbf{Y} - \mathbf{Y}\sum_{i=1}^{K} {\widehat{\Gamma}}_i {\widehat{\Gamma}}^{'}_i, 
\end{equation}
which represent the estimates of the specific factor ($\varepsilon$) realizations $\widehat{\mathbf{E}}= (e_{it})$. Thus  $\widehat e_i$ (or $\widehat e_j$), $i,j=1,...,p$ is a $1 \times T$ vector of one time series, while $\widehat e_t$, $t=1,...,T$ is a  $p \times 1$ vector of all time series at the one moment.

\subsubsection{Estimating blocks using $k$-means clustering}

Due to the heteroscedasticity of the idiosyncratic components, in the clustering procedure a correlation-based distance measure is used instead of the usual Euclidean distance:
\begin{equation}
d(\widehat{{e}}_i,\widehat{{e}}_j) = 1 - r_{ij},
\end{equation}
where $r_{ij}$ is the Pearson correlation coefficient between pairs of residual components $\widehat{{e}}_i$ and $\widehat{{e}}_j$, $i,j = 1,...,p$. 
The algorithm \cite{Alpaydin2020} minimizes the loss function 
\begin{equation}
\argmin_{b_1,..., b_p; \mathbf{\mu}_1,...,\mathbf{\mu}_M} \sum_{m = 1}^{M} \sum_{i = 1}^{p} b_i^{(m)} d(\widehat{{e}}_i,\widehat{\mu}_m),
\end{equation}
where $b_i^{(m)}$ is the binary indicator variable that assigns each data point to a cluster
\begin{equation}
b_i^{(m)} = \begin{cases}
    1, & m = \argmin_i d(\widehat{{e}}_i,\widehat{\mu}_m)  \\
    0, & \text{otherwise,} \\
\end{cases}
\end{equation}
and the centroid of a cluster is the average of the cluster members' residuals
\begin{equation}
    {\mu}_m = \dfrac{\sum_{i = 1}^{p} b_i^{(m)} {e}_i}{  \sum_{i = 1}^{p} b_i^{(m)} }.
\end{equation}
The algorithm is iterative and does not have a closed form solution. It converges to the local minimum, and depends on the initialization -- thus a repeated procedure with different initializations is used.

To determine the number of clusters from the data, we develop an iterative cross-validation procedure over the number of clusters $M$, where $M = 1,...,p$, as described in \ref{iterate}. Finally, the blocks of the block-diagonal idiosyncratic covariance are given by the clusters estimated using this procedure. 

We label this clustering-shrinkage estimator based on $k$-means \textit{CSK}, the corresponding idiosyncratic estimates $\widehat{\mathbf{\Psi}}_{CSK}$, and the entire covariance estimate $\widehat{\mathbf{\Sigma}}_{CSK}$.

\subsubsection{Estimating blocks using hierarchical clustering} 
As a more flexible framework, we also develop an estimator based on hierarchical clustering. Firstly, we propose a distance matrix based on the adaptive thresholding introduced by Cai and Liu \cite{Cai2011}, and specifically on the expression for the adaptive parameter from the formula (\ref{para}), and its implications for the hard thresholding rule. We can observe that the estimation of the final idiosyncratic covariance $\widehat{\mathbf{\Psi}}_{\tau_{ij}}$ depends on the relation of the full orthogonal complement entries $\widehat{S}_{ij}$ and the associated thresholding parameter $\tau_{ij}$, for $i,j = 1,...,p$. This means that if $|\widehat{S}_{ij}|< \tau_{ij}$ the final idiosyncratic component value is set to zero, otherwise if $|\widehat{S}_{ij}| \geq \tau_{ij}$ the value $\widehat{S}_{ij}$ remains unchanged. The relation in fact determines whether the two time series $i$ and $j$ are in the same cluster or not. Therefore, we use it as a custom similarity measure within the hierarchical clustering framework. Specifically, based on the relation, we define a distance matrix $\mathbf{D} = (D_{ij})$ which specify the dissimilarity of the two time series $i$ and $j$ as:
\begin{equation}
	{D}_{ij} =\begin{cases} \Bigg ( \dfrac{|\widehat{S}_{ij}|}{\sqrt{{\widehat{\theta}_{ij} T^{-1}{\log p}}}} \Bigg )^{-1}, & i \neq j \\
		0, & i = j, \end{cases}
\end{equation}
where $\widehat{\theta}_{ij} = T^{-1} \sum_{t=1}^{T} (e_{it} e_{jt} - \widehat{S}_{ij})^2$.

In order to evaluate different possible clusterings within a hierarchical framework, a linkage function $d(\cdot)$ is used, of which there are plenty: average linkage and weighted average linkage \cite{RobertReuvenSokal1958}, median and centroid linkage \cite{Sneath1973}, Ward linkage \cite{Ward1963}, single and complete linkage \cite{Alpaydin2020}. However, our focus is mainly on the methods less susceptible to noise, aligned with non-metric distance and forming the globular shape like average and weighted average. The considered linkage functions are given in detail in the Appendix \ref{beta}, and their detailed descriptions can be found in respective papers \cite{Sneath1973, Ward1963, RobertReuvenSokal1958, Alpaydin2020} -- the proposed hierarchical clustering estimator may rely on any of these.

Finally, we use an agglomerative clustering algorithm to build the clusters, based on the proposed distance matrix $\mathbf{D}$ and considered linkage functions $d(\cdot)$. We build the clustering tree and save each calculated distance (between points -- time series, and/or objects -- formed clusters) in the vector of distances $[L_1, L_2,...,Lp]$ each of which is related to a certain number of clusters $M$. To determine the optimal cutoff distance $L$, the estimator uses the iterative cross-validation procedure as described in \ref{iterate}. Finally, the blocks of the block-diagonal idiosyncratic covariance are given by the clusters estimated using this procedure. 

We label this clustering-shrinkage estimator based on the hierarchical clustering approach \textit{CSH}, the corresponding idiosyncratic estimates $\widehat{\mathbf{\Psi}}_{CSH}$, and the entire covariance estimate $\widehat{\mathbf{\Sigma}}_{CSH}$.

\subsubsection{Iterative procedure for selecting the number of blocks} 
\label{iterate}
The hyperparameters of the two clustering approaches also need to be estimated -- for {$k$-means clustering} this is directly the number of clusters, and for the {hierarchical clustering} algorithm this is the threshold at which the agglomerative tree is cut off. To obtain the values of these parameters we propose an $H$-fold cross-validation procedure, based on the residuals $\{ \mathbf{e}_{t} \}_{t \leq T}$. The residual series are split into a train subset $\{ \mathbf{e}_{t} \}_{t \in T_{train}}$ and a test subset $\{ \mathbf{e}_{t} \}_{t \in T_{test}}$, where $T_{train} + T_{test} = T$. The procedure is repeated $H$ times. In each fold $h \in H$ the following is performed:
\begin{itemize}
    \item Build the full orthogonal complement covariance matrix: ${\widehat{\mathbf{S}}}_{train-h}$ on train data and ${\widehat{\mathbf{S}}}_{test-h}$ on test data.
    \item Apply the proposed clustering algorithms to the residual series to obtain groups. When using {hierarchical clustering}, search through the grid of distances $L \in [L_1, L_2,..., Lp]$ (where each distance is connected to specific number of clusters M), and when using {$k$-means clustering}, search through the grid of number of clusters $M \in [1, 2,..., p]$.
    \item The indicator matrix $\mathbf{C}$ is obtained simply as: 
\begin{equation}
    C_{ij} = 
    \begin{cases}
		1  & i \text{ and } j \text{ are in the same cluster},\\
		0  & \mathrm{otherwise}.
	\end{cases}
\end{equation}
\item Calculate the validation error $Err_\varphi^h$, $h = 1,...H$ as the Frobenius norm of the difference between the cluster based idiosyncratic covariance estimate from the train set $\mathbf{\widehat{\Psi}}^{C}_{train-h}$ and the full orthogonal complement from the test set $\mathbf{\widehat{S}}_{test-h}$:
\begin{equation} \label{Err}
	Err^h_\varphi = \| \mathbf{\widehat{\Psi}}^{C}_{train-h}-  \mathbf{\widehat{S}}_{test-h}   \|_F^2.
\end{equation}

\end{itemize}  
We consider the mean of H validation errors assigned to each hyperparameter ($L$ in case of hierarchical and $M$ in the case of k-means clustering):
\begin{equation}
    Err^*_\varphi = \dfrac{1}{H} \sum_{h=1}^H Err^h_\varphi.
\end{equation}
and iterate over a grid of possible values of the considered hyperparameters (in the case of hierarchical clustering we iterate over a grid $\varphi = [L_1, L_2,..., L_p]$ and in case of $k$-means clustering we iterate over a grid $\varphi = [1, 2,..., p]$). The chosen distance criteria minimizes the error:
\begin{equation}
    \varphi = \argmin_\varphi Err^*_\varphi.
\end{equation}
For the hierarchical clustering the chosen hyperparameter is $\varphi_{CSH} = L_*$ and for the $k$-means clustering the chosen hyperparameter is $\varphi_{CSK} = M_*$. The final block-diagonal idiosyncratic covariance estimates ($\widehat{\mathbf{\Psi}}_{CSK}$ and $\widehat{\mathbf{\Psi}}_{CSH}$) are calculated on the entire estimation window using the selected hyperparameter.

\begin{figure}[H] 	
    \centering
    \begin{subfigure}[b]{0.49\textwidth}
         \centering
         \includegraphics[width=\textwidth]{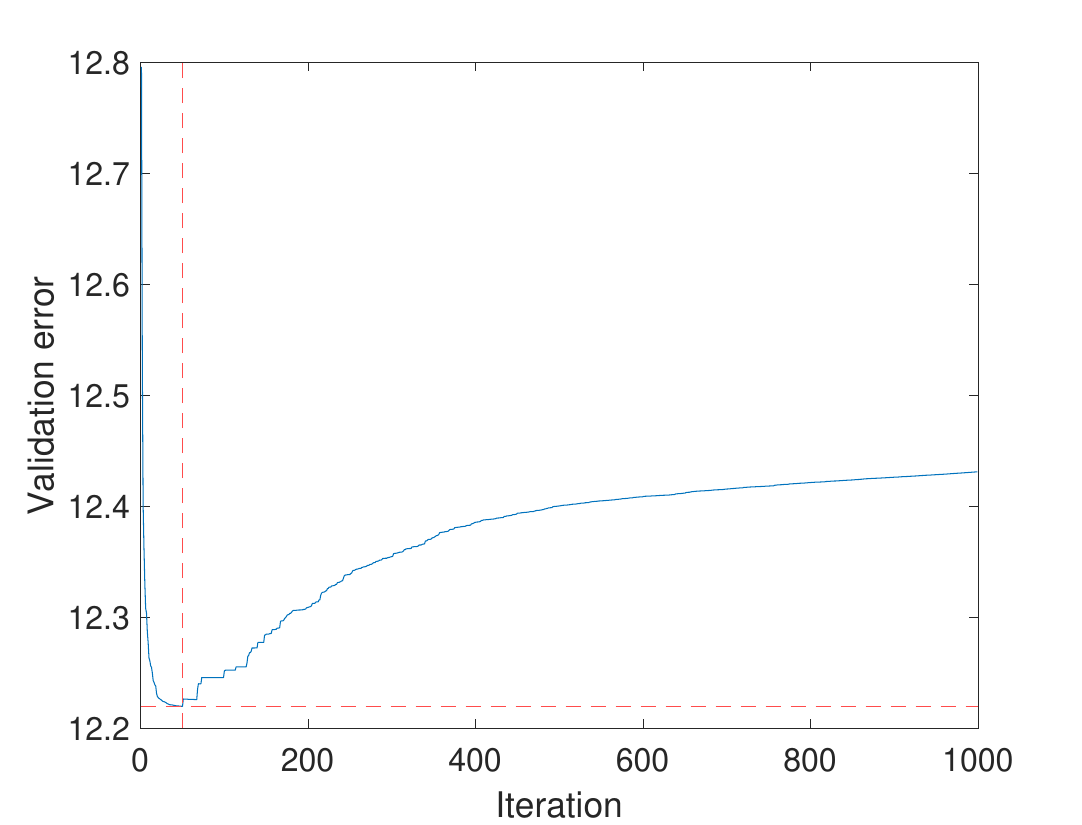}
         \caption{CSH estimator.}
         \label{fig:three sin x}
     \end{subfigure}
     \hfill
     \begin{subfigure}[b]{0.49\textwidth}
         \centering
         \includegraphics[width=\textwidth]{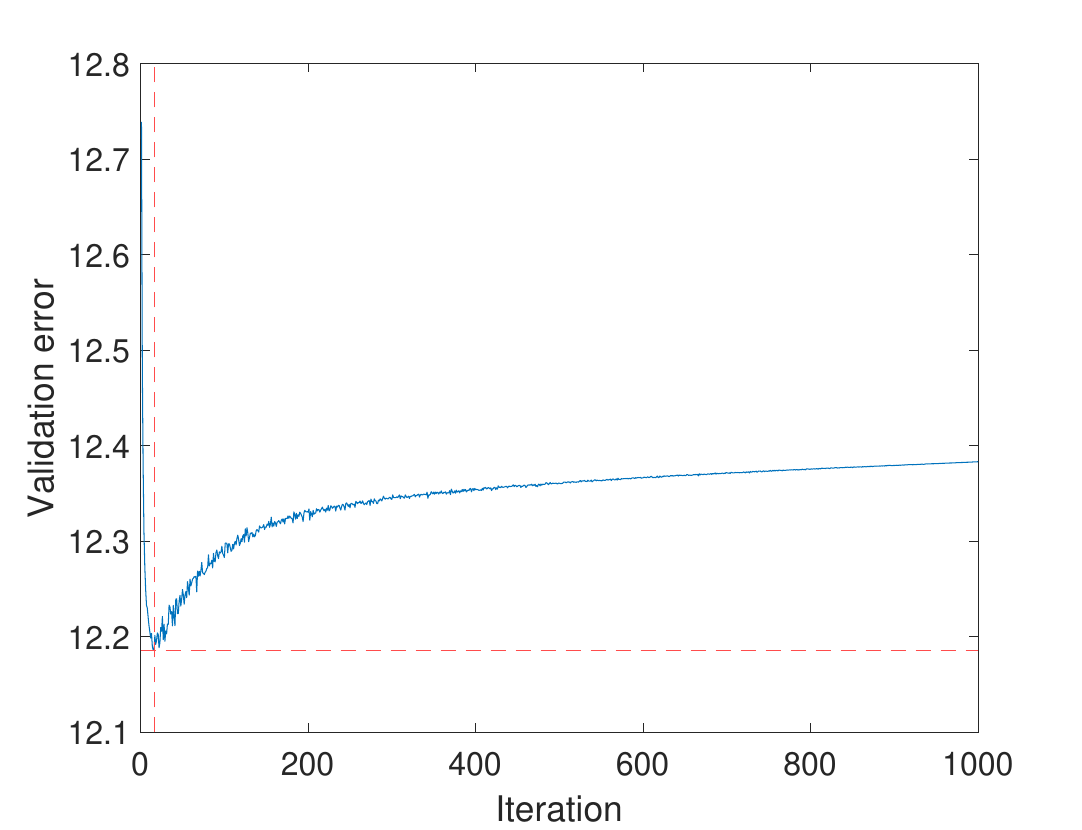}
         \caption{CSK estimator.}
         \label{fig:five over x}
     \end{subfigure}
        \caption{Validation errors through iterations over the hyperparameter space for both estimators performed on an example estimation window using historical market data. The red lines show the minimum value and the iteration it was reached in.}
        \label{fig:three graphs}
\end{figure}

Furthermore, we use the stability of the validation loss function to improve the algorithm and its speed. We use a stopping (convergence) criterion which stops the iteration loop when the change in the loss function is below a predefined threshold. Specifically, we monitor the average loss over 3 iterations and stop when this average stagnates. For the general shape of the cross-validation errors for the two approaches, see Figure \ref{fig:three graphs} which shows the cross-validation errors (blue line) through the iterations (over a grid of hyperparameter values) for the two considered estimators. It is evident that the validation errors exhibit an optimum which can be reached relatively quickly, without the need for traversing the entire hyperparameter space.

\section{Data and performance measures}
\subsection{Simulation data}
To test the ability of the estimator to identify true patterns of a block-diagonal structure, we construct a simulation scenario which allows us to analyze the performance of the estimators with respect to a known population covariance matrix. In the simulations we construct the common covariance $\mathbf{B}\mathrm{Cov}(F)\mathbf{B}'$ and the idiosyncratic covariance $\mathbf{\Psi}$ separately. The resulting covariance matrix is $\mathbf{\Sigma} = \mathbf{B}\mathrm{Cov}(F)\mathbf{B}' +\mathbf{\Psi}$.

\subsubsection{Generating the common component}
Without loss of generality, we assume that the factors have identity covariance, leaving all the variability to the factor loadings matrix $\mathbf{B}$. To simulate a random loadings matrix $\mathbf{B}$ we use the following procedure:
 \begin{enumerate}
     \item Generate random orthogonal loadings of unit length.
     \item Scale loadings so the first factor has average loading equal to 1 (this is in line with factor models in finance where the market factor is often the strongest and the loadings of assets towards this factor are centered around 1).
     \item Scale loadings by factor variances.
     \item Calculate the common covariance as $\mathbf{B}\mathbf{B}'$.
 \end{enumerate}

\subsubsection{Generating the idiosyncratic component}
We define data generating processes based on two different shapes of the idiosyncratic covariance. Firstly, a \textit{full block-diagonal} structure has a predetermined number of blocks of equal size and all series belong to one of the blocks. Secondly, a more difficult \textit{partial block-diagonal} structure does not use a predefined number of blocks and has a variable block size, thus allowing a large number of "blocks" with only a single series. In both cases, each cluster group is tapered and the correlations within are diminishing further from the diagonal. To construct the idiosyncratic covariance we first build the correlation matrix, which is then transformed into the covariance matrix. 

\noindent The full block-diagonal idiosyncratic covariance is constructed in the following way:
\begin{enumerate}
    \item Start from the identity correlation matrix $\mathbf{R} = \mathbf{I}$.
    \item For a predefined number of clusters $M$ generate uniform random cluster sizes.
    \item For each cluster add off-diagonal correlations following the tapering structure:
    \begin{equation}
      \mathbf{R}_{jk}  = {const} \cdot {base} ^ {({exponent} \cdot | j-k|)}
      \label{eq:tapering}
    \end{equation}
    \item Calculate the covariance matrix from the obtained correlation matrix and the idiosyncratic variances.
\end{enumerate} 

\noindent The partial block-diagonal covariance is constructed using graphs:
\begin{enumerate}
    \item Define a probability that a node (asset) is connected.
    \item Iterate over all assets -- for each asset, determine whether it will be connected (given the probability above) -- if yes, connect it to any one randomly selected asset (selected uniformly across the remaining $p-1$ assets).
    \item This procedure will build a graph with a number of connected components -- each component will correspond to a cluster, and ultimately, a block in the idiosyncratic covariance. Note that assets which are not connected remain as single-asset clusters.
    \item For each cluster larger than $1$ (i.e. other than single-asset clusters) add off-diagonal correlations following the tapering structure as described in \eqref{eq:tapering}.
    \item Calculate the covariance matrix from the obtained correlation matrix and the idiosyncratic variances.
\end{enumerate}

\subsection{Historical data}
We consider a collection of daily US stock returns from January 1995 to January 2017. The database consists of a large number of stocks, and at each time step we select the top $p$ stocks by market capitalization at the time defined by current date and considering only stocks which satisfy the following conditions:
\begin{enumerate}
    \item All marketcap and return data is available for the full training and test periods.
    \item There is at least 1 day of non-zero returns in test period and in train period.
    \item All the stocks have SIC sector identification.
\end{enumerate}
To determine the group membership in the \textit{CSI} estimator, we collect the Standard Industrial Classification Codes (SIC) sector codes for the selected stocks.

\subsection{Performance measures}
A most commonly used performance measure for determining the quality of matrix estimation is the Frobenius norm of the error: 
\begin{equation}
    \| \mathbf{\widehat{\Sigma}} - \mathbf{{\Sigma}} \|_F = \sqrt{\sum_{i=1}^{p} \sum_{j=1}^{T} |\hat\sigma_{ij} - \sigma_{ij}|^2}.
\end{equation}
However, it is generally a rough way to measure the covariance estimation quality. Since we focus on identifying the sparse correlation patterns in the idiosyncratic part, we can also focus on measuring how well these are identified by different estimators. In the simulation scenario, the population idiosyncratic covariance is known and thus we are able to measure the accuracy of identifying the true non-zero and zero elements in the population idiosyncratic covariance \cite{Zignic2022}.

We denote the classes of each element of the population idiosyncratic covariance with $0$ if the element is zero and $1$ if the element is non-zero. We use several, most common classification performance measures to evaluate the ability of the proposed estimators to identify the true sparsity patterns \cite{Sokolova2009}: positive rate (TP), true negative rate (TN), accuracy (Acc) and F1 score.
\begin{itemize}
    \item Accuracy is simply the ratio of correctly identified elements to the total number of off-diagonal elements in the idiosyncratic covariance.
    \item  TP (recall) and TN are the ratios of the correctly classified positive (negative) elements to the total number of positive (negative) elements in the population matrix.
    \item F1 is defined as the harmonic mean of \textit{recall} and \textit{precision}, where recall equals TP and precision is the ratio of classified true positives to the number of all predicted positives.
\end{itemize}

Moreover, we also consider clustering performance measures, since the population idiosyncratic covariance is considered to be block-diagonal. Rand's index measures the extent to which the obtained grouping corresponds to the reference grouping (for instance that of the population covariance). It is calculated as the accuracy of classification at the level of pairs of series, and is defined as follows:
\begin{equation}
    RI = \dfrac{TP+TN}{\binom{n}{2}},
\end{equation}
where ${\binom{n}{2}}$ represents the total number of possible pairs \cite{Rand1971, Warrens2022}.
We use the RI both in simulation data (with respect to the population idiosyncratic covariance) and historical data, where we use it to measure the similarity of different methods to the industry classification.

In addition to the results reported for the mentioned performance measures, which are averaged over a large number of simulations, for the simulation data we also consider the number of simulations in which the proposed \textit{CSH} and \textit{CSK} estimators outperform the benchmarks. Let $n_+$ denote the number of outcomes in which the considered estimator outperforms a benchmark for a given performance measure, in a total number of $n$ simulations. For the proportion $n_+/n$ we apply a non-parametric one sided paired sign test with following hypotheses: 
\begin{itemize}[label=]
 \item $H_0$: The probability of the estimator outperforming the given benchmark is 0.5. 
\item $H_1$: The probability is greater than 0.5.
\end{itemize}
Under the null hypothesis, $n_+$ follows a binomial distribution $B(n,0.5)$, which is directly used to calculate the corresponding $p$-value. We apply the test to each reported performance measure, in order to confirm whether the proposed approach achieves statistically significant improvements over the benchmark methods \cite{Begusic2020, Zignic2022}.

\subsection{Potfolio optimization} 
In addition to the performance measures defined above, we also consider a portfolio optimization scenario. We use minimum variance portfolios (in this context variance quantifies risk), since they highly depend on the quality of the estimated covariance -- the noise in the estimator indirectly transmits to the portfolio weights (variance minimizers are estimation-error maximizers \cite{Michaud1989}). The vector of portfolio weights $\boldsymbol{w}=[w_{1},...,w_{p}]'$ contains the percentage of the total capital allocated to each of the $p$ assets. When asset returns $\boldsymbol{Y}$ are arithmetic returns, the portfolio return is simply stated as $r_p = \boldsymbol{w}'\boldsymbol{Y}$. Then the portfolio variance is the variance of the linear combination $\sigma_p^2 = \mathbf{w'} {\boldsymbol{\Sigma}} \mathbf{w}$, which forms the basis for portfolio optimization in the mean-variance sense. The optimal portfolio weights for the minimum variance portfolio are then calculated by solving the following optimization problem:

\begin{equation}
		\min_\mathbf{w} \mathbf{w'} \widehat{\boldsymbol{\Sigma}} \mathbf{w}  \hspace{0.3cm} \text{	s.t.} \hspace{0.2cm}	\mathbf{1' w} = \mathbf{1}, 
\label{eq:gmv}\end{equation}
where $\widehat{\boldsymbol{\Sigma}}$ is the covariance estimate of the asset returns and the term $\mathbf{w'} \widehat{\boldsymbol{\Sigma} }\mathbf{w}$ is the portfolio variance. To evaluate different estimators we first obtain the optimal portfolio $\widehat{\mathbf{w}}$ on a given estimation window using a covariance estimate $\widehat{\boldsymbol{\Sigma}}$, then we calculate the out-of-sample portfolio risk, which we quantify as volatility (standard deviation of the returns) \cite{Choi2019}:
\begin{equation}
	\sigma_p : = \sqrt{\widehat{\mathbf{w}}'  \mathbf{\Sigma}  \widehat{\mathbf{w}}}.
 \label{eq:portfolio_risk}
\end{equation}

When using simulation data, the population covariance $\mathbf{\Sigma}$ is known. Portfolios are optimized using the estimates obtained from the generated time series, and "out-of-sample" portfolio risk is calculated by using the known population covariance 
$\mathbf{\Sigma}$ in the expression \eqref{eq:portfolio_risk}.
For historical data, the population covariance is unknown thus the sample estimates $\widehat{\mathbf{\Sigma}}_s$ from a future holding period are used in expression \eqref{eq:portfolio_risk} -- this corresponds to a backtesting approach where the optimized portfolios are held on a given future time period, and the realized risk of these portfolios is calculated. The daily portfolio volatility is annualized by multiplying with $\sqrt{252}$.

\section{Results}
For the simulation, we fix the number of factors to $K=5$, and simulate time series of length $T=250$ using the Student's $t$-distribution with $5$ degrees of freedom and zero mean, in order to replicate the heavy tailed property of asset returns. The simulations are repeated a total of $250$ times. For the full block-diagonal idiosyncratic structure we use $M=10$ clusters, and in the partial block-diagonal procedure the number of clusters is random and is a consequence of the random connections. For the correlation tapering within the clusters, we use $const = 0.3$, ${base} = 0.9$, ${exponent} = 0.1$ which result in similar correlation distributions as observed in historical data. Factor variances are set to $(0.25/([1,2,...,K]^{0.5})^2$. 

Firstly, in order to justify the choice of the linkage function in the hierarchical clustering method we evaluate five linkage functions (average, weighted average, Ward, centroid, medoid) on the partial block-diagonal simulation case with $p=1000$ series. The results are shown in Table \ref{linkage_evaluation} -- the F1, RI, Frobenius norm of the error and the portfolio risk are shown with respect to the known population covariance. The results show that the method based on the average linkage function outperforms in all of the considered aspects, thus in the rest of the paper we focus on the \textit{CSH} estimator based on average linkage.

\begin{table}[H]
		\centering
		\caption{The table shows main performance measures for partial block-diagonal simulation case. Comparison is made to assess impact of different linkage measures in hierarchical clustering (\textit{CSH} estimator).}
 \begin{tabu}[t]{l c c c c c c c c c c c c c c c c c }
 		\toprule
     Estimator  & F1 ($\%$) & RI ($\%$) & $\|\widehat{\mathbf{\Sigma}}-{\mathbf{\Sigma}}\|_F$ & $\sigma_p$ ($\%$) \\
			\midrule
			 {Average linkage} & \textbf{87.289} & \textbf{97.908} & \textbf{16.409}  & \textbf{3.482} \\
			 {Weighted linkage} & 83.865 & 97.466 & 16.414 &  3.486  \\
   	 {WWard's linkage} & 74.236  & 96.091 & 16.421  & 3.499 \\
    	 {Centroid linkage} & 76.023  & 96.259  & 16.420 & 3.496 \\
	 {Medoid linkage} & 73.968 & 96.087 & 16.426 & 3.503 \\
\midrule
		\end{tabu}
	\label{linkage_evaluation}
	\end{table}

We simulate the data with higher-dimensional $p=1000$ series and lower-dimensional $p=300$ series, using the same simulation parameters, in order to test the behavior of the estimators for different dimensionalities. Table \ref{sparse_measures_both} report the results for both considered dimensionalities and the two idiosyncratic covariance cases: partial block-diagonal and full block-diagonal. For all the measures, we report the p-values results of the one sided pair sign tests, based on the number of experiments in which the \textit{CSH} (for the {partial block-diagonal case}) and the \textit{CSK} (for the {full block-diagonal case}) estimators outperformed all the other methods.

For the partial block-diagonal case, the \textit{CSH} estimator is expected to outperform, which is confirmed in the results, and statistically significant (for higher-dimensional series in all cases and for lower-dimensional in most of the cases) -- this is evidently due to the fact that the hierarchical clustering approach can better accommodate single assets as clusters and generally the different cluster sizes, while the $k$-means approach is well-suited for compact, uniform-sized clusters. The true positive rate (TP) may be higher in some cases for the \textit{CSK} estimator as it tends to merge multiple small and single-asset clusters to one of a few larger clusters. Furthermore, as the \textit{CSH} estimator captures small and even single-asset clusters, missing some clusters has a higher impact on the true positive rate. In the {full block-diagonal} idiosyncratic covariance case, the \textit{CSK} estimator outperforms all the benchmark methods in all aspects, for both considered dimensionalities. The classification measures are close or equal to $100\%$ of accuracy. Moreover, even though the \textit{CSK} estimator is expected to outperform the \textit{CSH} estimator on the full block-diagonal case, the performance of the hierarchical approach remains comparatively high and the differences are not as large. The first reason for this is that the {full block-diagonal} case is much easier for both {clustering based} estimators. And the second reason lies in the fact that \textit{CSH} is more flexible and able to deal with both simulation cases very well, while the \textit{CSK} struggles in the partial block-diagonal case. 

It is important to note that for both the partial and full block-diagonal shapes, the lower-dimensional simulations for $p=300$ represent a drastically easier task for the $k$-means approach (\textit{CSK} estimator). This is due to the fact that the clusters in that case are naturally smaller, and the tapering effect of the off-diagonal correlations in the simulated population matrices is much weaker for smaller clusters (the correlations within those clusters are higher on average). In both considered dimensionalities, we observe that the {clustering based} estimators consistently outperform the {thresholding based estimators} -- however, in lower-dimensional case, the improvement is not as drastic and not as pervasive as in the higher-dimensional case. Evidently, the hierarchical approach benefits from the high dimensionality.

\begin{table}[H]
		\centering
		\caption{Table shows all the performance measures (rand index, classification measures, Frobenius norm) and out-of-sample portfolio volatility for the proposed estimators (\textit{CSH}, \textit{CSK}) and considered benchmark estimators. We reported two simulation cases {partial block-diagonal case} and {full block-diagonal case} for higher ($p=1000$) and lower ($p=300$) dimension. The p-values of the paired sign test comparing the \textit{CSH} (for partial block-diagonal case) and \textit{CSK} (for full block-diagonal case) estimator with all other methods are given in parentheses (below each method compared to the \textit{CSH} or \textit{CSK} estimator).}
 \begin{tabu}[t]{l l l c c c c c c c c }
 \toprule
 \toprule
 
 Case & Dim & Est  & F1 ($\%$) & Acc ($\%$) & TP ($\%$)& TN ($\%$) & RI ($\%$) & $\sigma_p$ ($\%$) & $\|\widehat{\mathbf{\Sigma}}-{\mathbf{\Sigma}}\|_F$ \\
\midrule
Partial & $1000$ &	{CSH} & \textbf{86.037} & \textbf{98.816} & \textbf{84.654 } & \textbf{99.579} & \textbf{97.691} & \textbf{3.465} & \textbf{16.653} \\
			
 & &   \textit{CSK} & 59.670 & 95.902 & 80.431 & {96.705} & 92.195 & {3.614} &{16.679}\\

    \rowfont{\scriptsize}  & &   & $(0)$ & $(0)$ & $(0.280)$ & $(0)$ & $(0)$ & $(0.004)$ & $(0.148)$  \vspace{5pt}\\
			
  & &     \textit{SCAD} &  41.997 & 92.368 & 76.078 & 93.206 & 85.935 &{7.093} & {16.700}   \\
    
    \rowfont{\scriptsize}  & &  &  $(0)$ & $(0)$ & $(0.096)$ & $(0)$ & $(0)$ &  $(0)$ & $(0.308)$  \vspace{5pt}\\
    
  & &     \textit{AL} &  60.814 & 97.001 & 63.126 & 98.571 & 94.211 & {7.994} & {16.706}   \\
    
    \rowfont{\scriptsize}& & &    $(0)$ & $(0)$ & $(0)$ & $(0.096)$ & $(0)$  &  $(0)$ & $(0.260)$\vspace{5pt}\\
    
   & &    \textit{SOFT} &  34.117 & 88.759 & 79.873 & 89.267 & 80.11 & {4.864} & {16.697}   \\
    
    \rowfont{\scriptsize}& & &   $(0)$ & $(0)$ & $(0.256)$ & $(0)$ & $(0)$ &  $(0)$ & $(0.340)$ \vspace{5pt}\\

\midrule

Partial &$300$ &	{CSH} & \textbf{78.163} & \textbf{94.081} & {78.698} & \textbf{96.903} & \textbf{89.666} & {6.186} & \textbf{4.993} \\
			
   & &  \textit{CSK} & {74.579} & {93.637} & \textbf{81.963} & 95.892 & {88.489} & \textbf{6.164} & {4.995}\\

    \rowfont{\scriptsize}  & && $(0.388)$ & $(0.328)$ & $(0.66)$ & $(0.244)$ & $(0.328)$ & $(0.416)$ & $(0.376)$  \vspace{5pt}\\
			
    & & \textit{SCAD} &  44.767 & 79.345 & 78.472 & 80.101 & 67.552 &{11.699} & {5.012}   \\
    
    \rowfont{\scriptsize}  & &&  $(0.028)$ & $(0)$ & $(0.472)$ & $(0.008)$ & $(0)$ &  $(0)$ & $(0.352)$  \vspace{5pt}\\
    
    & & \textit{AL}  &  55.002 & 87.965 & 68.560 & 91.335 & 79.119 & {12.717} & {5.015}   \\
    
   \rowfont{\scriptsize} & & &  $(0.044)$ & $(0.088)$ & $(0.176)$ & $(0.136)$ & $(0.088)$  &  $(0)$ & $(0.296)$\vspace{5pt}\\
    
   & &  \textit{SOFT} &  36.292 & 67.457 & 85.615 & 65.290 & 56.884 & {7.852} & {5.004}   \\   
   
    \rowfont{\scriptsize}&  & & $(0.028)$ & $(0)$ & $(0.556)$ & $(0)$ & $(0)$ &  $(0)$ & $(0.44)$ \vspace{5pt}\\

\midrule
\midrule

Full & 1000 & \textit{CSK} & \textbf{99.988} & \textbf{99.998} & \textbf{99.987} & \textbf{99.999} & \textbf{99.995} & \textbf{4.728} & \textbf{16.523} \\

  & & \textit{CSH} & 96.825 & 99.384 & 95.072 & 99.863 & 98.784 & {4.767} & {16.534}\\
						
\rowfont{\scriptsize}   & & & $(0.008)$ & $(0.008)$ & $(0.004)$ & $(0.016)$ & $(0.008)$ & $(0.12)$ & $(0.228)$\vspace{5pt}\\

  & & \textit{SCAD} & 61.451 & 91.658 & 66.46 & 94.458 & 84.725 &{8.425} &{16.686} \\

\rowfont{\scriptsize}&  & &  $(0)$ & $(0)$ & $(0)$ & $(0)$ & $(0)$ & $(0)$ & $(0.292)$ \vspace{5pt}\\

  & & \textit{AL} &  59.131 & 93.828 & 44.705 & 99.286 & 88.429 & {10.088} & {16.726} \\
			
\rowfont{\scriptsize}&  & &  $(0)$ & $(0)$ & $(0)$ & $(0.01)$ & $(0)$ & $(0)$ & $(0.264)$ \vspace{5pt}\\

  & & \textit{SOFT} &  55.592 & 88.269 & 73.407 & 89.921 & 79.312 & {8.928} & {16.660}   \\
			
\rowfont{\scriptsize}   & & & $(0)$ & $(0)$ & $(0)$ & $(0)$ & $(0)$ & $(0)$ & $(0.312)$ \vspace{5pt}\\

\midrule



Full & 300 & \textit{CSK} & \textbf{99.998} & \textbf{1.000} & \textbf{99.998} & \textbf{1.000} & \textbf{99.999} & \textbf{6.641} & \textbf{5.025} \\

  & & \textit{CSH} & 96.337 & 99.173 & 99.050 & 99.187 & 98.401 & {6.724} & {5.028}\\
						
\rowfont{\scriptsize}  & &  & $(0)$ & $(0)$ & $(0)$ & $(0)$ & $(0)$ & $(0.008)$ & $(0.196)$\vspace{5pt}\\

  & & \textit{SCAD} & 51.816 & 81.976 & 96.817 & 80.330 & 70.554 &{13.449} &{5.060} \\

\rowfont{\scriptsize}  & & & $(0)$ & $(0)$ & $(0)$ & $(0)$ & $(0)$ & $(0)$ & $(0.320)$ \vspace{5pt}\\

  & & \textit{AL} &  63.917 & 89.375 & 94.030 & 88.860 & 81.076 & {13.987} & {5.057} \\
			
\rowfont{\scriptsize}  & & & $(0)$ & $(0)$ & $(0)$ & $(0)$ & $(0)$ & $(0)$ & $(0.332)$ \vspace{5pt}\\

  & & \textit{SOFT} &  37.307 & 66.825 & 98.721 & 63.284 & 55.770 & {8.303} & {5.054}   \\
			
\rowfont{\scriptsize}  & &  & $(0)$ & $(0)$ & $(0)$ & $(0)$ & $(0)$ & $(0)$ & $(0.4)$ \vspace{5pt}\\   

\bottomrule
   
		\end{tabu}
		\label{sparse_measures_both}
	\end{table}

\begin{figure}[H] 	
    \centering
    \begin{subfigure}[b]{0.47\textwidth}
         \centering
         \includegraphics[width=\textwidth]{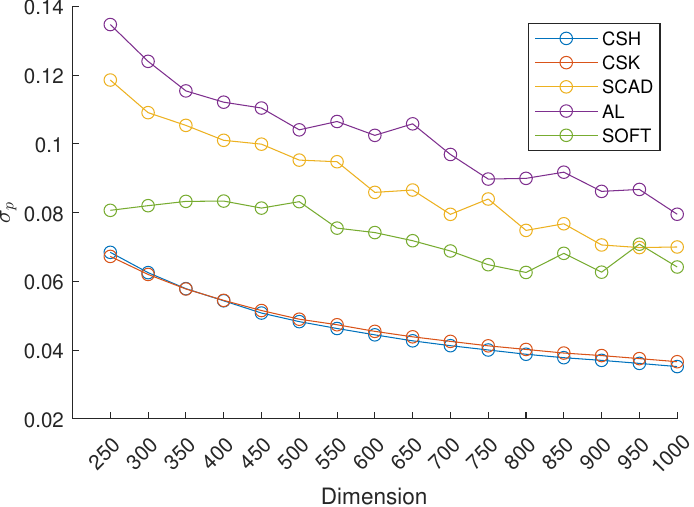}
         \caption{Partial block-diagonal case.}
     \end{subfigure}
     \hfill
     \begin{subfigure}[b]{0.47\textwidth}
         \centering
         \includegraphics[width=\textwidth]{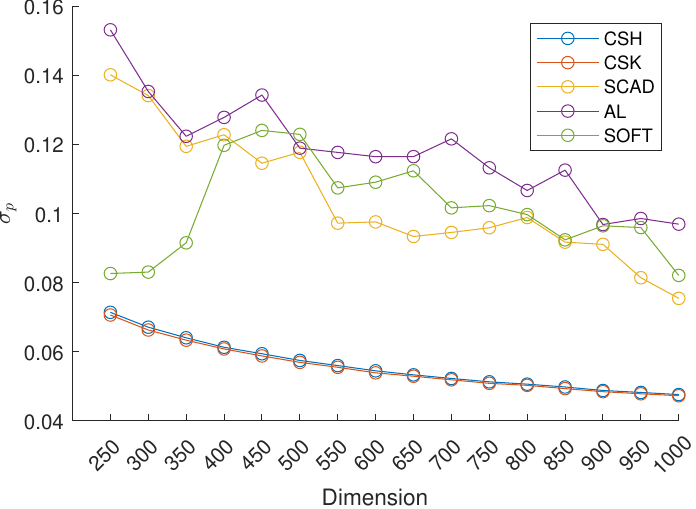}
         \caption{Full block-diagonal case.}
     \end{subfigure}
			\caption{Out-of-sample portfolio risk $\sigma_p$ for different dimensionalities of the data. The sample window length is $T=250$ and the dimension varies from $p=250$ to $p=1000$ with a step size of $50$.}
        \label{fig:portfolio_risk_dimensionality}
\end{figure}

Although we did not observe drastic structural differences, or differences in order of the estimators' performances, there are some differences in the behavior of the estimators when increasing the dimension. In figure \ref{fig:portfolio_risk_dimensionality} we show the average out-of-sample portfolio risks over all simulations for the {partial block-diagonal case} and the {full block-diagonal} case over different dimensions $p$, starting from $p=250$ to $p=1000$ with a step of $50$. Evidently, the clustering based estimators benefit widely from increased dimensionality in both simulation scenarios, and consistently outperform the thresholding based estimators over all dimensions. Moreover, the {clustering based} estimators also show a great level of stability as clustering procedures are much more robust in capturing the idiosyncratic groupings, while the thresholding based estimators only focus on high pairwise covariances, and might miss out on the other members of the groups -- these properties of the estimators are also the reason behind the F1 and RI being much higher for {clustering based methods} in Table \ref{sparse_measures_both} for both shown dimensions.

\begin{figure*}[h!] 
    \begin{minipage}{\textwidth}
    \centering
    \includegraphics[scale=0.77]{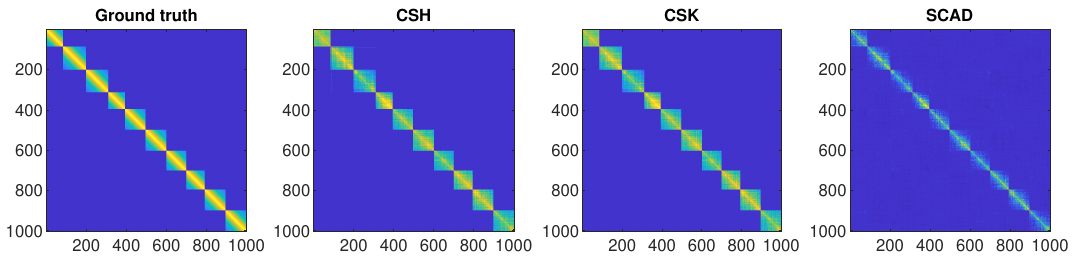} 
    \includegraphics[scale=0.77]{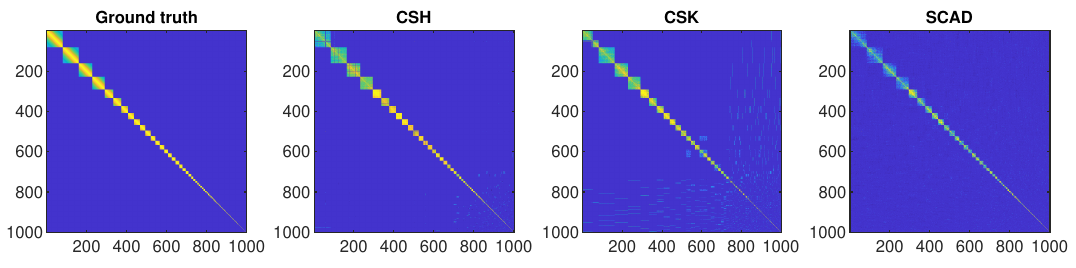} 
    \caption{Plot of the simulated idiosyncratic covariance (ground truth) for a single simulation case, in comparison to the idiosyncratic covariance estimated by the \textit{CSH}, \textit{CSK} and the \textit{SCAD} estimators. Top row shows the full block-diagonal case, and the bottom row shows the partial block-diagonal case. Blue areas on the matrices correspond to zero-valued entries.}
       \label{fig:blocks}
    \end{minipage}
\end{figure*}

To demonstrate visually how different estimators work, we show how each estimator forms the idiosyncratic covariances for both simulation cases in Figure \ref{fig:blocks}. In addition to the two clustering based estimators we also show the \textit{SCAD} estimator, since it performs the best out of the thresholding based approaches. The full block-diagonal case seems to be captured almost perfectly by both clustering based methods, while the \textit{SCAD} methods evidently misses out on some elements with smaller correlations (due to tapering further away from the diagonal). For the  {partial block-diagonal} case larger clusters tends to be broken into smaller ones by the \textit{CSK} approach, and some small and single-asset clusters are combined into larger ones. The \textit{SCAD} approach identifies the smaller ones but mostly misses out on the larger ones. The \textit{CSH} estimator is shown to capture the clusters relatively well.

\subsection{Historical data results}
\begin{figure}[H]
    \begin{minipage}{\textwidth}
    \centering
    \includegraphics[width=0.87\textwidth]{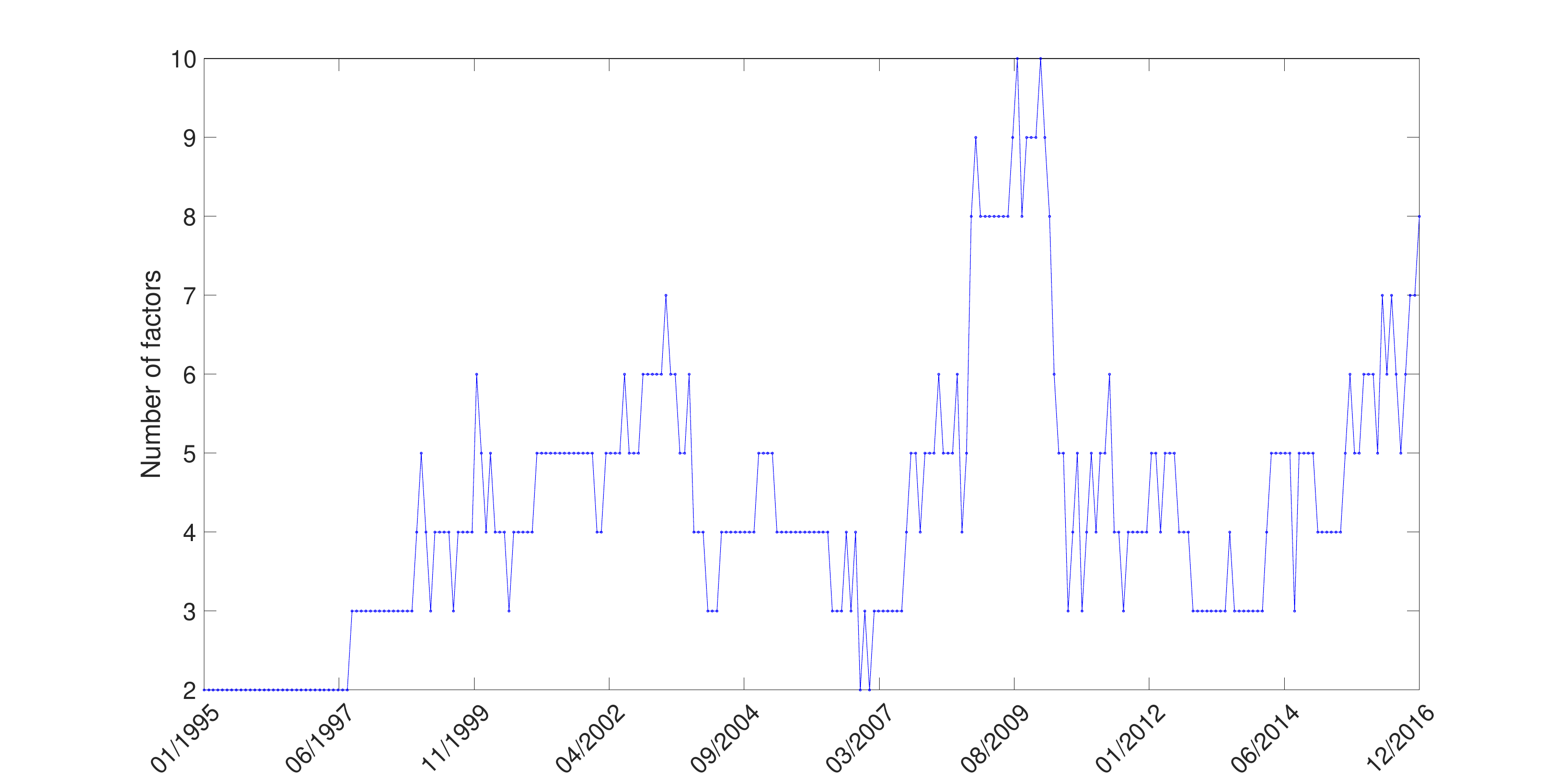} 
    \caption{Estimated number of factors throughout the historical time period, for $p=1000$ assets.}
    \label{fig:num_factors}
    \end{minipage}
\end{figure}

We test the estimators on the historical data using a portfolio optimization approach -- at each time step, the portfolios are constructed using the covariance estimated during the past $1$ year of daily returns (a total of $252$ data points) using the considered estimators. The portfolio is held for the next month ($22$ days) and the portfolio volatility is calculated on this out-of-sample future holding period. This approach assumes that the covariances estimated in the past $1$-year window continue to hold on the future $1$-month period, and the future returns are considered as realizations of this process. The total number of iterations is $264$. In historical data we are letting the algorithm to find the number of common factors $\widehat{K}$. To estimate the number of factors within each time window we use the \textit{Bai-Ng IC1} method. Figure \ref{fig:num_factors} shows the evolution of the estimated number of common factors through time -- the number of factors ranges from $2$ to $10$ with the average of $4.33$ and the median and mode being the same and equal to $4$. 

As in historical data we have no access to the ground truth, we can only observe the performance in terms of the portfolios optimized using the different estimators. We also calculate the Rand index to observe the similarity of the methods to the industry grouping. This similarity is not something we wish to maximize, but rather an information about how different estimators behave in relation to the groups given by the industry grouping. We perform the analysis for two different numbers of assets: $p=1000$ and $p=300$, by choosing the $p$ stocks with the highest market capitalization at each time step. The results are shown in Table \ref{table:results_hist}.

\begin{table}[H] 
		\centering
		\caption{Portfolio volatility $\sigma_p$ and the average RI (similarity to industries defined in SIC) over the historical testing period for the different estimators on historical data, shown for $p=300$ and $p=1000$.}
 \begin{tabu}[t]{r | c c | c c }
 		\toprule
   & \multicolumn{2}{c|}{$p=1000$} & \multicolumn{2}{c}{$p=300$}\\
     Estimator &  $\sigma_p$ (\%)  & RI (\%) & $\sigma_p$ (\%)  & RI (\%)\\
			\midrule
			\textit{CSH}  & 6.553&  72.467   & 8.666&  70.165  \\
                \textit{CSK}   & \textbf{6.324} &   65.953  & {8.599} &   60.855  \\
                 \textit{CSI} &   6.382&   100.00 & \textbf{8.548}&   100.00\\
                \textit{SCAD}   &   8.958&  72.405  & 10.488&  62.131 \\
                \textit{AL}   & 10.669& 73.584 & 10.060& 68.441 \\
                 \textit{SOFT}   & 8.670& 69.718  &  8.575& 59.751 \\
			\bottomrule
		\end{tabu}
   \label{table:results_hist}
\end{table}

The results show that the {clustering based} methods generally outperform the {thresholding based} estimators, with the exception of the \textit{SOFT} thresholding estimator for the lower dimensional case. The clustering based estimators manage to find clusters which are not so similar to the industry classifications, as suggested by the RI results -- yet these alternative groupings seem to perform similarly or even better than the industry classifications. This result affirms the hypothesis that the industry groupings may not be optimal, depending on the application. While still performing better than {thresholding based} estimators in lower-dimensional cases, we see that {clustering based} estimators benefit drastically from the increased dimensionality. Nevertheless, the results also suggest that the industry classification is in fact a valuable contributor to the performance -- \textit{CSI} shows excellent performance. However, the industry classification data may not always be available, depending on the asset universe or different markets one might consider. On the other hand, the proposed clustering based approach only requires historical return data.

\begin{figure}[h]
    \begin{minipage}{\textwidth}
        \centering
        \includegraphics[scale=0.82]{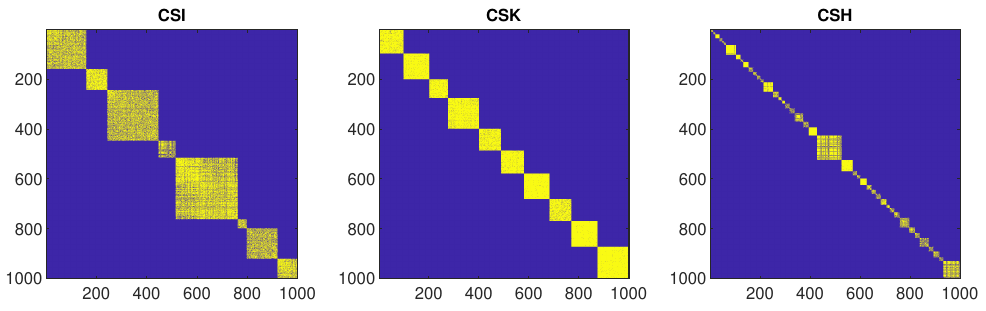} 
        \caption{Block-diagonal idiosyncratic covariances obtained by using respective \textit{CSI}, \textit{CSK} and \textit{CSH} estimators. The blue areas represent the zero-valued entries.}
        \label{fig:block_picuture_real}
    \end{minipage}
\end{figure}

Finally, we also inspect the shapes of the identified idiosyncratic covariances for a specific time window in the historical data. Figure \ref{fig:block_picuture_real} shows the idiosyncratic covariance given different methods: \textit{CSI}, \textit{CSK} and \textit{CSH}. It is important to note that for each method, the assets were sorted according to the corresponding clustering results (so that the assets in the same clusters are placed next to each other). The differences are quite visible -- the \textit{CSI} features relatively big blocks of varying sizes, and different values of off-diagonal elements, while the \textit{CSK} finds more smaller and compact groups. The estimated idiosyncratic covariance using the \textit{CSH} estimator differs mostly from the other methods. It allows small one-member clusters but does not omit the relevant information (bigger clusters are also observed). Due to this flexibility, the average number of clusters is much larger than the number given by the \textit{CSK} estimator, and the matrix is generally more sparse.

\section{Conclusion}
We consider the problem of estimating the covariance matrix of high-dimensional financial return time series, given an underlying latent factor model. The latent factor model allows for a specific structure of the covariance matrix -- a low-rank component due to common factors and a full-rank sparse idiosyncratic component. In this paper we specifically focus on the estimation of the idiosyncratic component under the assumption that the considered financial assets form groups, even after accounting for common factors, which has recently been documented in the literature. This leads to a block-diagonal structure of the idiosyncratic covariance. We follow a two step estimation procedure where the first step consists of estimating the common component, and in the second step the residual component is used to obtain a sparse estimate of the idiosyncratic covariance. We formulate a unified approach to estimating the block-diagonal idiosyncratic covariance and consider several methods to obtain the unknown block structure (clusters). We also propose an iterative cross-validation procedure in the context of the squared error given the assumed latent factor model, and test the proposed approach on simulation data and historical return data.

The simulation results show that the proposed clustering based estimators successfully recognize the true sparse idiosyncratic covariance patterns, while decreasing the optimized portfolio volatility. Moreover, they show other desirable properties: both clustering approaches benefit from increased dimensionality and demonstrate stable results for different numbers of simulated series. The hierarchical approach implemented in the CSH estimator shows great versatility, since it is able to capture the difficult patterns given by the partial block-diagonal idiosyncratic case, while retaining performance for the full-block diagonal case. Tests on historical data confirmed the superiority of the {clustering based} estimators with respect to the thresholding based estimators. A striking finding is that the groups identified by the proposed estimators seem to differ to a great extent from the industry classification, however the portfolio performance of the proposed estimators is on par with or even better than the industry based estimator CSI.

The results evidently affirm the basic research hypothesis of the paper -- that estimating the sparse idiosyncratic covariance as a block-diagonal matrix improves upon the thresholding based approach. Allowing the assets to form entire clusters dramatically enriches the space of idiosyncratic covariance estimates, and ultimately results in a more realistic model of the asset return dependence. The proposed approach will hopefully make its way to applications in risk modeling of high-dimensional return time series in broad asset universes, and especially those where an industry or similar classification is not known a priori. We also hope to inspire new research, especially in the area of modeling the hierarchical group structures and their effect on the idiosyncratic covariance in approximate factor models.

\section*{Funding}
This work was supported in part by the Croatian Science Foundation under Project 5241.

\bibliography{BIB14.bib}

\begin{thebibliography}{10}

\bibitem{LedoitPower2022}
O.~Ledoit and M.~Wolf, ``{The Power of (Non-)Linear Shrinking: A Review and Guide to Covariance Matrix Estimation},'' {\em Journal of Financial Econometrics}, vol.~20, pp.~187--218, jan 2022.

\bibitem{Lam2020b}
C.~Lam, ``{High-dimensional covariance matrix estimation},'' {\em Wiley Interdisciplinary Reviews: Computational Statistics}, vol.~12, pp.~1--21, oct 2020.

\bibitem{Choi2019}
Y.~G. Choi, J.~Lim, and S.~Choi, ``{High-dimensional Markowitz portfolio optimization problem: empirical comparison of covariance matrix estimators},'' {\em Journal of Statistical Computation and Simulation}, vol.~89, pp.~1278--1300, may 2019.

\bibitem{Bun2017}
J.~Bun, J.~P. Bouchaud, and M.~Potters, ``{Cleaning large correlation matrices: Tools from Random Matrix Theory},'' {\em Physics Reports}, vol.~666, pp.~1--109, jan 2017.

\bibitem{MohsenPourahmadi2013}
{Mohsen Pourahmadi}, {\em {High-Dimensional Covariance Estimation}}.
\newblock Wiley Series in Probability and Statistics, Hoboken, NJ, USA: John Wiley \& Sons, Inc., jun 2013.

\bibitem{Goldberg2022}
L.~R. Goldberg and A.~N. Kercheval, ``{James–Stein for the leading eigenvector},'' {\em Proceedings of the National Academy of Sciences of the United States of America}, vol.~120, no.~2, 2023.

\bibitem{Ross1976}
S.~A. Ross, ``{The arbitrage theory of capital asset pricing},'' {\em Journal of Economic Theory}, vol.~13, pp.~341--360, dec 1976.

\bibitem{Fama1993}
E.~F. Fama and K.~R. French, ``{Common risk factors in the returns on stocks and bonds},'' {\em Journal of Financial Economics}, vol.~33, pp.~3--56, feb 1993.

\bibitem{Fama2015}
E.~F. Fama and K.~R. French, ``{A five-factor asset pricing model},'' {\em Journal of Financial Economics}, vol.~116, pp.~1--22, aug 2015.

\bibitem{Connor1995}
G.~Connor, ``{The Three Types of Factor Models: A Comparison of Their Explanatory Power},'' {\em Financial Analysts Journal}, vol.~51, pp.~42--46, may 1995.

\bibitem{Fan2011a}
J.~Fan, Y.~Liao, and M.~Mincheva, ``{High-dimensional covariance matrix estimation in approximate factor models},'' {\em Annals of Statistics}, vol.~39, pp.~3320--3356, nov 2011.

\bibitem{Fan2013}
J.~Fan, Y.~Liao, and M.~Mincheva, ``{Large covariance estimation by thresholding principal orthogonal complements},'' {\em Journal of the Royal Statistical Society. Series B: Statistical Methodology}, vol.~75, pp.~603--680, aug 2013.

\bibitem{Lettau2020}
M.~Lettau and M.~Pelger, ``{Estimating latent asset-pricing factors},'' {\em Journal of Econometrics}, vol.~218, pp.~1--31, sep 2020.

\bibitem{Fan2016}
J.~Fan, Y.~Liao, and H.~Liu, ``{An overview of the estimation of large covariance and precision matrices},'' {\em The Econometrics Journal}, vol.~19, pp.~C1--C32, feb 2016.

\bibitem{Begusic2019}
S.~Begu{\v{s}}i{\'{c}} and Z.~Kostanj{\v{c}}ar, ``{Cluster-Based Shrinkage of Correlation Matrices for Portfolio Optimization},'' in {\em 11th International Symposium on Image and Signal Processing and Analysis (ISPA 2019)}, pp.~301--305, IEEE, sep 2019.

\bibitem{Fan2008}
J.~Fan, Y.~Fan, and J.~Lv, ``{High dimensional covariance matrix estimation using a factor model},'' {\em Journal of Econometrics}, vol.~147, pp.~186--197, nov 2008.

\bibitem{Chamberlain1983}
G.~Chamberlain and M.~Rothschild, ``{Arbitrage, Factor Structure, and Mean-Variance Analysis on Large Asset Markets},'' {\em Econometrica}, vol.~51, p.~1281, sep 1983.

\bibitem{Bickel2008}
P.~J. Bickel and E.~Levina, ``{Covariance regularization by thresholding},'' {\em The Annals of Statistics}, pp.~2577--2604, 2008.

\bibitem{Rothman2009}
A.~J. Rothman, E.~Levina, and J.~Zhu, ``{Generalized thresholding of large covariance matrices},'' {\em Journal of the American Statistical Association}, vol.~104, pp.~177--186, mar 2009.

\bibitem{Cai2011}
T.~Cai and W.~Liu, ``{Adaptive thresholding for sparse covariance matrix estimation},'' {\em Journal of the American Statistical Association}, vol.~106, pp.~672--684, jun 2011.

\bibitem{Wang2017}
W.~Wang and J.~Fan, ``{Asymptotics of empirical eigenstructure for high dimensional spiked covariance},'' {\em Annals of Statistics}, vol.~45, pp.~1342--1374, jun 2017.

\bibitem{Ait-Sahalia2017}
Y.~A{\"{i}}t-Sahalia and D.~Xiu, ``{Using principal component analysis to estimate a high dimensional factor model with high-frequency data},'' {\em Journal of Econometrics}, vol.~201, pp.~384--399, dec 2017.

\bibitem{Begusic2020}
S.~Begu{\v{s}}i{\'{c}} and Z.~Kostanj{\v{c}}ar, ``{Cluster-Specific Latent Factor Estimation in High-Dimensional Financial Time Series},'' {\em IEEE Access}, vol.~8, pp.~164365--164379, sep 2020.

\bibitem{Zignic2022}
L.~Zignic, S.~Begusic, and Z.~Kostanjcar, ``{Estimating the Block-Diagonal Idiosyncratic Covariance in High-Dimensional Factor Models},'' in {\em 2022 International Conference on Software, Telecommunications and Computer Networks (SoftCOM)}, pp.~1--6, oct 2022.

\bibitem{Ando2017}
T.~Ando and J.~Bai, ``{Clustering Huge Number of Financial Time Series: A Panel Data Approach With High-Dimensional Predictors and Factor Structures},'' {\em Journal of the American Statistical Association}, vol.~112, no.~519, pp.~1182--1198, 2017.

\bibitem{Dorfleitner2003}
G.~Dorfleitner, ``{Why the return notion matters},'' {\em International Journal of Theoretical and Applied Finance}, vol.~6, pp.~73--86, nov 2003.

\bibitem{Bai2019}
J.~Bai and S.~Ng, ``{Rank regularized estimation of approximate factor models},'' {\em Journal of Econometrics}, vol.~212, no.~1, pp.~78--96, 2019.

\bibitem{Bai2002}
J.~Bai and S.~Ng, ``{Determining the number of factors in approximate factor models},'' {\em Econometrica}, vol.~70, pp.~191--221, jan 2002.

\bibitem{Zou2006}
H.~Zou, ``{The adaptive lasso and its oracle properties},'' {\em Journal of the American Statistical Association}, vol.~101, pp.~1418--1429, dec 2006.

\bibitem{Fan2001}
J.~Fan and R.~Li, ``{Variable selection via nonconcave penalized likelihood and its oracle properties},'' {\em Journal of the American Statistical Association}, vol.~96, pp.~1348--1360, dec 2001.

\bibitem{Fan2016a}
J.~Fan, A.~Furger, and D.~Xiu, ``{Incorporating Global Industrial Classification Standard Into Portfolio Allocation: A Simple Factor-Based Large Covariance Matrix Estimator With High-Frequency Data},'' {\em Journal of Business and Economic Statistics}, vol.~34, pp.~489--503, oct 2016.

\bibitem{Ledoit2004}
O.~Ledoit and M.~Wolf, ``{A well-conditioned estimator for large-dimensional covariance matrices},'' {\em Journal of Multivariate Analysis}, vol.~88, pp.~365--411, feb 2004.

\bibitem{Ledoit2004a}
O.~Ledoit and M.~Wolf, ``{Honey, I shrunk the sample covariance matrix, The Journal of Portfolio Management},'' {\em The Journal of Portfolio Management}, vol.~30, pp.~110--119, jul 2004.

\bibitem{Alpaydin2020}
E.~Alpaydin, {\em {Introduction to Machine Learning}}.
\newblock Cambridge, MA: MIT Press, mar 2020.

\bibitem{RobertReuvenSokal1958}
C.~D.~M. {Robert Reuven Sokal}, ``{A Statistical Method for Evaluating Systematic Relationships},'' {\em University of Kansas Science Bulletin}, vol.~62, pp.~1902--1996, mar 1958.

\bibitem{Sneath1973}
P.~H.~A. Sneath and R.~R. Sokal, {\em {Numerical Taxonomy: The Principles and Practice of Numerical Classification}}.
\newblock W. H. Freeman, second~ed., jan 1973.

\bibitem{Ward1963}
J.~H. Ward, ``{Hierarchical Grouping to Optimize an Objective Function},'' {\em Journal of the American Statistical Association}, vol.~58, pp.~236--244, jun 1963.

\bibitem{Sokolova2009}
M.~Sokolova and G.~Lapalme, ``{A systematic analysis of performance measures for classification tasks},'' {\em Information Processing {\&} Management}, vol.~45, pp.~427--437, jul 2009.

\bibitem{Rand1971}
W.~M. Rand, ``{Objective Criteria for the Evaluation of Clustering Methods},'' {\em Journal of the American Statistical Association}, vol.~66, p.~846, dec 1971.

\bibitem{Warrens2022}
M.~J. Warrens and H.~van~der Hoef, ``{Understanding the Adjusted Rand Index and Other Partition Comparison Indices Based on Counting Object Pairs},'' {\em Journal of Classification}, vol.~39, pp.~487--509, nov 2022.

\bibitem{Michaud1989}
R.~O. Michaud, ``{The Markowitz Optimization Enigma: Is ‘Optimized' Optimal?},'' {\em Financial Analysts Journal}, vol.~45, pp.~31--42, jan 1989.

\end{thebibliography}
\bibliographystyle{ieeetr}

\clearpage

\appendix

\section{Linkage functions}
Denote with $G$ the cluster group set in the hierarchy which contains the observations. We describe five main types in detail:
\label{beta}
\begin{itemize}

\item \textbf{Average linkage} \cite{RobertReuvenSokal1958} is average inter-cluster distance calculated as distance between each pair of the observations in each cluster.
\begin{equation}
D_{G_1 G_2} = d_{avg} (G_1, G_2) = \dfrac{1}{T_1 T_2} \sum_{{\widehat{e}}_i \in G_1} \sum_{{\widehat{e}}_j \in G_2} d({\widehat{e}}_i, {\widehat{e}}_j)
\end{equation}

\item \textbf{Centroid linkage} \cite{Sneath1973} is the distance between the centroids of the two clusters. 
\begin{equation}
D_{G_1 G_2} = d_{cen} (G_1, G_2) = \Big \| \dfrac{1}{T_1} \sum_{{\widehat{e}}_i \in G_1}  {\widehat{e}}_i -  \dfrac{1}{T_2} \sum_{{\widehat{e}}_j \in G_2}  {\widehat{e}}_j \Big \|_2
\end{equation}

\item \textbf{Median linkage} \cite{Sneath1973} is Euclidean distance between weighted centroids of the two clusters.
\begin{equation}
D_{G_1 G_2} = d_{med} (G_1, G_2) = \Big \| {\tilde{e}}_{1} - {\tilde{e}}_{2} \Big \|_2
\end{equation}
where ${\tilde{e}}_{1}$ and ${\tilde{e}}_{2}$ are weighted centroids of clusters $G_1$ and $G_2$. If the cluster $G_1$ is created by combining two clusters $G_{1a}$ and $G_{1b}$, then ${\tilde{e}}_{1} = \dfrac{1}{2}({\tilde{e}}_{1a} + {\tilde{e}}_{1b})$. 

\item \textbf{Ward distance} \cite{Ward1963} is defined as the within-cluster sum of the squares of the distances between all objects in the cluster and the centroid of the cluster.
\begin{align}
D_{G_1 G_2} = d_{ward} (G_1, G_2) = \notag \\ 
\sqrt{\dfrac{2 \cdot T_1 T_2}{(T_1+T_2)}}& \Big \| \dfrac{1}{T_1} \sum_{{\widehat{e}} \in G_1}  {\widehat{e}} -  \dfrac{1}{T_2} \sum_{{\widehat{e}}\in G_2}  {\widehat{e}}\Big \|_2
\end{align}

\item \textbf{Weighted average linkage} \cite{RobertReuvenSokal1958} is defined recursively. If cluster $G_1$ is created by combining clusters $G_{1a}$ and $G_{1b}$ than the distance between the cluster $G_1$ and $G_2$ is defined as average of the distance between $G_{1a}$ and $G_2$ and the distance between $G_{1b}$ and $G_2$.  
\end{itemize}

\section{Shrinkage intensity} \label{alpha}
We outline established Ledoit and Wolf procedure for the optimal shrinkage intensity \cite{Ledoit2004a}. 
The optimal shrinkage intensity $\alpha_m$, should minimize the expected value of the quadratic loss function
\begin{equation}
	P(\alpha_m) =	\| \alpha_m \widehat{\mathbf{S}}^{C_m} + (1-\alpha_m) \mathbf{\Tilde{S}}^{C_m} - \mathbf{S}^{C_m} \|^2,
\end{equation}
where $\mathbf{S}^{C_m}$ is the unknown population covariance.

The optimal $\alpha_m$ estimate from the Ledoit and Wolf procedure \cite{Ledoit2004} for shrinkage estimator, is given as
\begin{equation}
	\widehat{\alpha}^*_m = \max \Bigg \{ 0, \min \Bigg \{ \dfrac{\widehat{\kappa}^m}{T_m} , 1 \Bigg \} \Bigg \},
\end{equation}
where 
\begin{equation}
	\widehat{\kappa}^m = \dfrac{\widehat{\pi}^m-\widehat{\rho}^m}{\widehat{\gamma}^m}.
\end{equation}
and $T_m$ is the block size.

For the simplicity we will drop prefix $m$, but all the following parts are calculated per each block. The constant term $\widehat{\pi}$ is consistent estimator of asymptotic variances of the sample block matrix entries $\mathbf{\widehat{S}}^{C}$ scaled by $\sqrt{T}$ (size of the block) defined as \cite{Ledoit2004} :
\begin{equation}
    \widehat{\pi} = \dfrac{1}{T} \sum_{i=1}^{N} \sum_{i=j}^N \Big (e_{it} - \bar{e}_i) (e_{jt} - \bar{e}_j) - \widehat{S}_{ij}    \Big )^2
\end{equation}
where $\bar{e}_i$ is the sample average of the returns of stock $i$ from the cluster block $c$.
Term $\widehat{\rho}$ is consistent estimator of sum of asymptotic covariances of the shrinkage target entries with the block sample covariance entries scaled by $\sqrt{T}$ (size of the block) defined as:

\begin{equation}
\widehat{\rho} = \sum_{i=1}^N \widehat{\pi}_{ii} + \sum_{i=1}^N \sum_{j=1, j \neq i}^N \dfrac{\tilde{r}}{2} \Bigg ( \sqrt{\dfrac{\widehat{S}_{jj}}{\widehat{S}_{ii}}} \widehat{\eta}_{ii,ij} + \dfrac{\widehat{S}_{ii}}{\widehat{S}^m_{jj}} \widehat{\eta}_{jj,ij}    \Bigg )
\end{equation}
where 
\begin{align}
 &   \widehat{\eta}_{ii,ij} = \dfrac{1}{T_m} \Big \{ (e_{it} - \bar{e}_i)^2 - \widehat{S}_{ii}   \Big \} \times \Big \{  (e_{it} - \bar{e}_i) (e_{jt} - \bar{e}_j) - \widehat{S}_{ij}   \Big \}, \\
 &    \widehat{\eta}_{jj,ij} = \dfrac{1}{T_m} \Big \{ (e_{jt} - \bar{e}_j)^2 - \widehat{S}_{jj}   \Big \} \times \Big \{  (e_{it} - \bar{e}_i) (e_{jt} - \bar{e}_j) - \widehat{S}_{ij}   \Big \} 
\end{align}

And $\widehat{\gamma}$ is a consistent estimator of misspecification of the (population) shrinkage target defined as:
\begin{equation}
 \widehat{\gamma} = \sum_{i=1}^{N} \sum_{i=j}^N (\bar{r} \sqrt{\widehat{S}_{ii} \widehat{S}_{jj}})^2
\end{equation}

\end{document}